\documentclass[10pt,conference]{IEEEtran}
\usepackage{cite}
\usepackage{amsmath,amssymb,amsfonts}
\usepackage{algorithmic}
\usepackage{graphicx}
\usepackage{textcomp}
\usepackage{xcolor}
\usepackage{hyperref}
\usepackage{booktabs}
\usepackage{nicematrix}
\usepackage{multirow}

\newcommand{\grayrowshort}[1]{\cellcolor{gray!20}{#1-5,#1-6}}
\newcommand{\eat}[1]{}

\def\BibTeX{{\rm B\kern-.05em{\sc i\kern-.025em b}\kern-.08em
    T\kern-.1667em\lower.7ex\hbox{E}\kern-.125emX}}
\begin{document}

\title{An Empirical Comparison of Pre-Trained Models of Source Code}



\author{
\IEEEauthorblockN{Changan Niu\IEEEauthorrefmark{1},
Chuanyi Li\IEEEauthorrefmark{1},
Vincent Ng\IEEEauthorrefmark{2},
Dongxiao Chen\IEEEauthorrefmark{1},
Jidong Ge\IEEEauthorrefmark{1},
Bin Luo\IEEEauthorrefmark{1}}
\IEEEauthorblockA{\IEEEauthorrefmark{1}State Key Laboratory for Novel Software Technology, Nanjing University, Nanjing, China\\
Email: niu.ca@outlook.com, lcy@nju.edu.cn, MF21320014@smail.nju.edu.cn, {gjd,luobin}@nju.edu.cn}
\IEEEauthorblockA{\IEEEauthorrefmark{2}Human Language Technology Research Institute, University of Texas at Dallas, Richardson, Texas, USA\\
Email: vince@hlt.utdallas.edu}
}

\maketitle

\begin{abstract}
While a large number of pre-trained models of source code have been successfully developed and applied to a variety of software engineering (SE) tasks in recent years, our understanding of these pre-trained models is arguably fairly limited. With the goal of advancing our understanding of these models, we perform the first systematic empirical comparison of 19 recently-developed pre-trained models of source code on 13 SE tasks. To gain additional insights into these models, we adopt a recently-developed 4-dimensional categorization of pre-trained models, and subsequently investigate whether there are correlations between different categories of pre-trained models and their performances on different SE tasks.
\end{abstract}

\begin{IEEEkeywords}
Pre-training of Source Code, AI for SE
\end{IEEEkeywords}

\section{Introduction}
\label{section:introduction}

Despite the successful application of deep learning to various Artificial Intelligence (AI) subfields such as natural language processing (NLP) and computer vision in recent years, a large amount of annotated training data is typically needed to train the millions or even billions of network parameters in a deep neural model. For many learning tasks, including those in software engineering (SE), obtaining annotated data is costly. To address this data annotation bottleneck, NLP researchers have come up with an idea that can arguably be considered one of the most exciting developments in recent deep learning research, namely {\em pre-training} \cite{dai:nips15,howard:acl18,peters2018deep,radford2018improving}. Rather than training a model from scratch (i.e., with randomly initialized network weights), which typically requires a lot of task-specific annotated data, one can first pre-train it on one or more so-called self-supervised tasks (i.e., tasks for which annotated data can be automatically generated and therefore large amounts of training data are readily available) so that its weights encode general linguistic and commonsense knowledge about language, and then the resulting pre-trained model can be fine-tuned to learn the target task using (a potentially small amount of) task-specific annotated training data in the usual supervised manner. A large number of pre-trained models of natural language (PTM-NLs) have been developed and widely used in NLP, such as BERT~\cite{devlin2019bert}, XLNet~\cite{yang2019xlnet}, RoBERTa~\cite{liu2019roberta}, ELECTRA~\cite{clark2019electra}, GPT-2~\cite{radford2019gpt2}, T5~\cite{raffel2020t5}, and BART~\cite{lewis2020bart}.

Soon thereafter, pre-trained models have made their way into SE research. Initial applications of pre-trained models in SE have primarily involved retraining PTM-NLs on source code~\cite{kanade2020cubert,buratti2020c-bert,svyatkovskiy2020gpt-c,de2021javabert,drain2021deepdebug}. Nevertheless, employing the resulting re-trained models (henceforth PTM-Cs) for SE tasks is not ideal, as there are code-specific characteristics that may not be properly taken into account by these models, such as the syntactic~\cite{alon2019code2vec,zhang2019novel} and semantic structures~\cite{ben2018neural} inherent in source code~\cite{karmakar2021pre}. Consequently, SE researchers have developed a number of pre-trained models of source code (henceforth CodePTMs) that take into account code-specific characteristics in the past few years~\cite{allamanis2018learning,cummins2020deep,hoang2020cc2vec,ma2022graphcode2vec,bui2021infercode,zhang2022learning}.

Despite the fact that a large number of CodePTMs have been successfully developed and applied to a variety of SE tasks in recent years, our understanding of CodePTMs is arguably fairly limited. Currently, only one survey of pre-trained models of source code is available from Niu et al.~\cite{niu2022deep}, but it just performs a summary and analysis from the results reported by the origin model. While pre-trained models are task-agnostic and therefore can be applied to different SE tasks by design, virtually all CodePTMs have been evaluated on only a handful of SE tasks. For instance, TreeBERT~\cite{jiang2021treebert}, has only been evaluated on code summarization and method name generation. This is by no means ideal: without knowing how TreeBERT performs on the remaining SE tasks, we do not know whether it can achieve state-of-the-art results on any of those tasks. This in turn implies that our understanding of these models could be partial and that the current state-of-the-art could have been very different had we evaluated the existing models on most, if not all, of the available SE tasks. Even when two pre-trained models are being evaluated on the same SE task, a head-to-head comparison of these models could still be made complicated if they are evaluated on different datasets available for this task~\cite{niu2022sptcode}.

With the goal of advancing our understanding of existing pre-trained models of source code, we conduct the first systematic empirical comparison of 19 recently-developed CodePTMs on 13 popular SE tasks. To gain additional insights into these CodePTMs, we employ a recently-developed four-dimensional categorization of CodePTMs~\cite{niu2022deep} to categorize existing the 19 CodePTMs used in our study, and subsequently investigate whether there are correlations between categories of CodePTMs and their performances on SE tasks.

\section{Experimental Setup}
\label{section:experimental_setup}

\subsection{SE Tasks}
\label{section:setup_se_tasks}

\begin{table}[t]
    \centering
    \caption{Details of evaluation tasks, datasets and metrics. }
    \label{table:se_tasks}
    \resizebox{\linewidth}{!}{%
        \begin{NiceTabular}{llllll}
        \CodeBefore
            \grayrowshort{2}
            \grayrowshort{7}
            \grayrowshort{9}
            \grayrowshort{10}
            \grayrowshort{12}
            \grayrowshort{14}
            \grayrowshort{16}
            \grayrowshort{19}
            \grayrowshort{20}
            \grayrowshort{24}
            \grayrowshort{25}
            \grayrowshort{26}
            \grayrowshort{33}
        \Body
            \toprule
                \textbf{Type} &
                \textbf{I-O} &
                \textbf{Task} &
                \textbf{Ab.} &
                \textbf{\texttt{ID}: Dataset} &
                \textbf{Metrics} \\
            \midrule
                \Block{14-1}{Und.} &
                \Block{8-1}{C-V} &
                \Block{5-1}{Defect Detection} &
                \Block{5-1}{DD} &
                \texttt{D1}: Devign~\cite{zhou2019devign} &
                Acc \\
            \cline{5-6}
                &
                &
                &
                &
                \texttt{D2}: DeepBugs~\cite{pradel2018deepbugs} &
                Acc \\
            \cline{5-6}
                &
                &
                &
                &
                \texttt{V1}: VMC~\cite{kanade2020cubert} &
                Acc \\
            \cline{5-6}
                &
                &
                &
                &
                \texttt{W1}: WBO~\cite{kanade2020cubert} &
                Acc \\
            \cline{5-6}
                &
                &
                &
                &
                \texttt{S3}: SO~\cite{kanade2020cubert} &
                Acc \\
            \cline{3-6}
                &
                &
                \Block{2-1}{Clone Detection} &
                \Block{2-1}{CD} &
                \texttt{B1}: BigCloneBench~\cite{svajlenko2014bigclonebench} &
                F1 \\
            \cline{5-6}
                &
                &
                &
                &
                \texttt{C2}: CLCDSA~\cite{nafi2019clcdsa} &
                F1 \\
            \cline{3-6}
                &
                &
                Exception Type &
                ET &
                \texttt{K1}: Kanade et al.~\cite{kanade2020cubert} &
                Acc \\
            \cline{2-6}
                &
                \Block{2-1}{C-C} &
                \Block{2-1}{Code-to-Code\\ Retrieval} &
                \Block{2-1}{CR} &
                \texttt{P1}: POJ-104~\cite{mou2016poj104} &
                MAP \\
            \cline{5-6}
                &
                &
                &
                &
                \texttt{C2}: CLCDSA~\cite{nafi2019clcdsa} &
                MRR \\
            \cline{2-6}
                &
                \Block{4-1}{NL-C} &
                \Block{2-1}{Code Search} &
                \Block{2-1}{CS} &
                \texttt{C3}: CodeSearchNet (Filtered)~\cite{lu2021codexglue} &
                MRR \\
            \cline{5-6}
                &
                &
                &
                &
                \texttt{A1}: AdvTest~\cite{lu2021codexglue} &
                MRR \\
            \cline{3-6}
                &
                &
                \Block{2-1}{Code Question\\ Answering} &
                \Block{2-1}{QA} &
                \texttt{C4}: CoSQA~\cite{huang2021cosqa}, WebQueryTest~\cite{lu2021codexglue} &
                MRR \\
            \cline{5-6}
                &
                &
                &
                &
                \texttt{F1}: FDM~\cite{kanade2020cubert} &
                Acc \\
            \hline
                \Block{18-1}{Gen.} &
                \Block{10-1}{C-C} &
                \Block{3-1}{Code Translation} &
                \Block{3-1}{CT} &
                \texttt{C5}: CodeTrans~\cite{lu2021codexglue} &
                EM/B./C.B. \\
            \cline{5-6}
                &
                &
                &
                &
                \texttt{T1}: TransCoder~\cite{roziere2020transcoder} &
                CA \\
            \cline{5-6}
                &
                &
                &
                &
                \texttt{C2}: CLCDSA~\cite{nafi2019clcdsa} &
                R.L \\
            \cline{3-6}
                &
                &
                \Block{1-1}{Bug Fixing} &
                \Block{1-1}{BF} &
                \texttt{B2}: BFP~\cite{tufano2019bfp} &
                EM/B./C.B. \\
            \cline{3-6}
                &
                &
                \Block{4-1}{Code Completion} &
                \Block{4-1}{CC} &
                \texttt{P2}: PY150~\cite{raychev2016py150} &
                EM/ES \\
            \cline{5-6}
                &
                &
                &
                &
                \texttt{C6}: CugLM~\cite{liu2020cuglm} &
                EM \\
            \cline{5-6}
                &
                &
                &
                &
                \texttt{S1}: SLM~\cite{alon2020anycodecompletion} &
                EM \\
            \cline{5-6}
                &
                &
                &
                &
                \texttt{S2}: Svyatkovskiy et al.~\cite{svyatkovskiy2020gpt-c} &
                PPL \\
            \cline{3-6}
                &
                &
                Mutant Generation &
                MG &
                \texttt{G1}: GM~\cite{tufano2019learning} &
                EM/B. \\
            \cline{3-6}
                &
                &
                \Block{1-1}{Assert Generation} &
                \Block{1-1}{AG} &
                \texttt{A3}: ATLAS~\cite{watson2020assert} &
                EM/B. \\
            \cline{2-6}
                &
                \Block{7-1}{C-NL} &
                \Block{7-1}{Code Summarization} &
                \Block{7-1}{SM} &
                \texttt{C3}: CodeSearchNet (Filtered)~\cite{lu2021codexglue} &
                B. \\
            \cline{5-6}
                &
                &
                &
                &
                \texttt{A2}: Attn2FC~\cite{haque2020improved} &
                B. \\
            \cline{5-6}
                &
                &
                &
                &
                \texttt{D3}: DeepCom~\cite{hu2018deepcom} &
                B. \\
            \cline{5-6}
                &
                &
                &
                &
                \texttt{P2}: PY150~\cite{raychev2016py150} &
                EM \\
            \cline{5-6}
                &
                &
                &
                &
                \texttt{C7}: code2seq~\cite{alon2019code2seq} &
                EM \\
            \cline{5-6}
                &
                &
                &
                &
                \texttt{T2}: TL-CodeSum~\cite{hu2018jcsd} &
                B. \\
            \cline{5-6}
                &
                &
                &
                &
                \texttt{M1}: Miceli-Barone and Sennrich~\cite{miceli2017pcsd} &
                B. \\
            \cline{2-6}
                &
                NL-C &
                Code Generation &
                CG &
                \texttt{C8}: CONCODE~\cite{iyer2018concode} &
                EM/B./C.B. \\
            \bottomrule
        \end{NiceTabular}
    }
\end{table}

Table~\ref{table:se_tasks} enumerates the 13 SE tasks we will use in our comparative experiments. These are also the SE tasks that are typically used to evaluate pre-trained models of source code. Following previous work~\cite{niu2022deep}, in the first two columns, we classify each task along two dimensions: (1) whether the task concerns \textit{understanding} (\textbf{Und.}) or \textit{generation} (\textbf{Gen.}); and (2) the type of input assumed by the task and the type of produced output (\textbf{I-O}), where \textbf{C}, \textbf{NL}, and \textbf{V} denote code, natural language, and extracted/predicted value, respectively. Table~\ref{table:se_tasks} also shows the abbreviation (Ab.), the dataset, and the main evaluation metrics for each task.

To make the number of experiments manageable in our comparison, whenever there are multiple datasets for a task, we choose the most popular one (shown in Gray in Table~\ref{table:se_tasks}) except for Code Search, where we chose \texttt{A1} over \texttt{C3} since \texttt{A1} is the filtered version of \texttt{C3} and the results on \texttt{A1} is more reflective of the generalization ability of a model~\cite{lu2021codexglue}.

\subsection{Evaluation Metrics}
\label{section:setup_metrics}

For each SE task, we will perform evaluations using the standard metrics listed in the last column of Table~\ref{table:se_tasks}. For classification and retrieval tasks, metrics such as Acc (Accuracy), F1, Precision (P), Recall (R), Mean Reciprocal Rank (MRR) and Mean Average Precision (MAP) are used. For generation tasks, metrics developed in the NLP community such as perplexity (PPL), Levenshtein edit similarity (ES)~\cite{svyatkovskiy2020gpt-c}, BLEU (B.)~\cite{papineni2002bleu}, as well as variants developed in the SE community, such as CodeBLEU (C.B.)~\cite{ren2020codebleu}, are used. Moreover, some generation tasks have also used variants of Accuracy for evaluation, one of which indicates whether the sequence generated by the model exactly matches (EM) the correct answer, and the other, Computational Accuracy (CA), computes the number of times the hypothesis function generates the same output as the reference when given the same inputs~\cite{roziere2020transcoder}.

\subsection{Pre-trained Models}
\label{section:setup_codeptms}

In this subsection, we first present an overview of 26 of the PTMs that have been applied to SE tasks and then enumerate the 19 pre-trained models of source code that we will include in our empirical comparison. 

\subsubsection{Categorization}
\label{section:setup_codeptms_categorization}

\begin{table*}[t!]
	\centering
	\caption{Categorization of existing pre-trained models and their performance on SE tasks as reported in their original papers. The strongest result for each dataset is boldfaced.}
	\label{table:codeptms}
	\resizebox{\textwidth}{!}{%
	\begin{NiceTabular}{lllllccc|c|cc|ccccc|c|c}
	\CodeBefore
	    \rowcolors{4}{gray!15}{}
	\Body
	    \toprule
            \Block{3-1}{\textbf{Models}} &
            \Block{3-1}{\textbf{Arch.}} &
            \Block{3-1}{\textbf{Modality}} &
            \Block{3-1}{\textbf{Tasks}} &
            \Block{3-1}{\textbf{PL}} &
            \Block{1-6}{\textbf{Code Understanding Tasks}} &
            &
            &
            &
            &
            &
            \Block{1-7}{\textbf{Code Generation Tasks}} &
            &
            &
            &
            &
            &
            \\
        \cline{6-18}
            &
            &
            &
            &
            &
            \Block{1-3}{\textbf{C-V}} &
            &
            &
            \Block{}{\textbf{C-C}} &
            \Block{1-2}{\textbf{NL-C}} &
            &
            \Block{1-5}{\textbf{C-C}} &
            &
            &
            &
            &
            \textbf{C-NL} &
            \textbf{NL-C} \\
        \cline{6-18}
            &
            &
            &
            &
            &
            \textbf{DD} &
            \textbf{CD} &
            \textbf{ET} &
            \textbf{CR} &
            \textbf{CS} &
            \textbf{QA} &
            \textbf{CT} &
            \textbf{BF} &
            \textbf{CC} &
            \textbf{MG} &
            \textbf{AG} &
            \textbf{SM} &
            \textbf{CG} \\
        \midrule
            RoBERTa~\cite{liu2019roberta} &
            TE &
            NL Text &
            MLM &
            - &
            D1:61.05 &
            B1:94.9 &
            &
            P1:76.67 &
            \Block{}{C3:61.7\\ A1:18.33} &
            C4:60.3 &
            &
            &
            &
            &
            &
            C3:16.57 &
            \\
        GPT-2~\cite{radford2019gpt2} &
            TD &
            NL Text &
            ULM &
            - &
            &
            &
            &
            &
            &
            &
            &
            &
            P2:41.73 &
            &
            &
            &
            17.35 \\
        BART~\cite{lewis2020bart} &
            TF &
            NL Text &
            DAE &
            - &
            &
            &
            &
            &
            &
            &
            &
            11.7 &
            &
            &
            &
            &
            \\
        T5~\cite{raffel2020t5} &
            TF &
            NL Text &
            Seq2Seq MLM &
            - &
            D1:61.93 &
            &
            &
            &
            &
            &
            &
            9.7 &
            &
            &
            &
            C3:18.35 &
            18.65 \\
        \hline
            SCELMo~\cite{karampatsis2020scelmo} &
            LSTM &
            Code &
            BiLM &
            Mono &
            D2:93.12 &
            &
            &
            &
            &
            &
            &
            &
            &
            &
            &
            &
            \\
        CuBERT~\cite{kanade2020cubert} &
            TE &
            Code &
            MLM + NSP &
            Mono &
            \Block{}{V1:95.21\\ W1:92.46\\ S3:93.36} &
            &
            79.12 &
            &
            &
            F1:98.09 &
            &
            &
            &
            &
            &
            \Block{}{P2:33.48\\ C7:52.76\\ D3:17.41} &
            \\
        GPT-C~\cite{svyatkovskiy2020gpt-c} &
            TD &
            Code &
            ULM &
            Multi &
            &
            &
            &
            &
            &
            &
            &
            &
            S2:1.65 &
            &
            &
            &
            \\
        C-BERT~\cite{buratti2020c-bert} &
            TE &
            Code &
            MLM &
            Mono &
            D1:57.4 &
            &
            &
            &
            &
            &
            &
            &
            &
            &
            &
            &
            \\
        JavaBERT~\cite{de2021javabert} &
            TE &
            Code &
            MLM &
            Mono &
            &
            &
            &
            &
            &
            &
            &
            &
            &
            &
            &
            &
            \\
        CodeGPT-adapted~\cite{lu2021codexglue} &
            TD &
            Code &
            ULM & 
            Multi &
            &
            &
            &
            &
            &
            &
            &
            &
            P2:42.37 &
            &
            &
            &
            20.1 \\
        DeepDebug~\cite{drain2021deepdebug} &
            TF &
            Code &
            Seq2Seq MLM &
            Mono &
            &
            &
            &
            &
            &
            &
            &
            15.05 &
            &
            &
            &
            &
            \\
        \hline
            CodeBERT~\cite{feng2020codebert} &
            TE &
            Code + Doc &
            MLM \& RTD &
            Multi &
            &
            &
            &
            P1:82.67 &
            C3:69.3 &
            C4:65.7 &
            C5:58.5 &
            10.8 &
            &
            &
            &
            \Block{}{C3:17.83\\ P2:35.97\\ C7:56.52\\ D3:17.87} &
            \\
        GraphCodeBERT~\cite{guo2021graphcodebert} &
            TE &
            \Block{}{Code + Doc +\\ DFG Nodes} &
            \Block{}{MLM +\\ EP + NA} &
            Multi &
            &
            B1:97.1 &
            &
            P1:85.16 &
            C3:71.3 &
            C4:68.4 &
            C5:59.1 &
            13.2 &
            &
            &
            &
            &
            \\
        CugLM~\cite{liu2020cuglm} &
            TE &
            Code &
            \Block{}{IMLM + NSP + ULM} &
            Multi &
            &
            &
            &
            &
            &
            &
            &
            &
            C6:81.91 &
            &
            &
            &
            \\
        DOBF~\cite{roziere2021dobf} &
            TF &
            Code &
            \Block{}{MLM \&\\ Seq2Seq IMLM} &
            Multi &
            &
            B1:95.9 &
            &
            &
            A1:38.3 &
            &
            T1:46.35 &
            &
            &
            &
            &
            C3:18.65 &
            \\
        T5-learning~\cite{mastropaolo2021t5-learning} &
            TF &
            Code \& Doc &
            Seq2Seq MLM &
            Mono &
            &
            &
            &
            &
            &
            &
            &
            6.5 &
            &
            28 &
            40.5 &
            A2:15 &
            \\
        PLBART~\cite{ahmad2021plbart} &
            TF &
            Code \& Post &
            DAE &
            Multi &
            D1:63.18 &
            B1:97.2 &
            &
            &
            &
            C4:65.0 &
            C5:64.8 &
            14.10 &
            &
            &
            &
            C3:18.32 &
            18.75 \\
        ProphetNet-Code~\cite{qi2021prophetnet-code} &
            TF &
            Code \& Doc &
            FNP &
            Multi &
            &
            &
            &
            &
            &
            &
            &
            &
            &
            &
            &
            C3:18.54 &
            \\
        CoTexT~\cite{phan2021cotext} &
            TF &
            Code + Doc &
            Seq2Seq MLM &
            Multi &
            D1:\textbf{65.99} &
            &
            &
            &
            &
            &
            &
            17.30 &
            &
            &
            &
            C3:18.38 &
            20.1 \\
        TreeBERT~\cite{jiang2021treebert} &
            TF &
            \Block{}{Code +\\ AST Paths} &
            TMLM + NOP &
            Multi &
            &
            &
            &
            &
            &
            &
            &
            &
            &
            &
            &
            \Block{}{\textbf{D3:20.49}\\ \textbf{P2:45.81}\\ \textbf{C7:67.9}} &
            \\
        OSCAR~\cite{peng2021oscar} &
            TE &
            IR + AEI &
            MLM + CCL &
            Mono &
            &
            &
            &
            P1:49.17 &
            &
            &
            &
            &
            &
            &
            &
            &
            \\
        CodeDisen~\cite{zhang2021codedisen} &
            LSTM &
            \Block{}{Code +\\ AST Seq} &
            \Block{}{VGVAE + CLR\\ + PD + ACP} &
            Multi &
            &
            C2:90.0 &
            &
            C2:43.6 &
            &
            &
            C2:50.08 &
            &
            &
            &
            &
            &
            \\
        CodeT5~\cite{wang2021codet5} &
            TF &
            Code + Doc &
            \Block{}{Seq2Seq MLM / IT /\\ Seq2Seq IMLM / BDG} &
            Multi &
            D1:65.78 &
            B1:97.2 &
            &
            &
            &
            C4:67.8 &
            C5:\textbf{66.4} &
            \textbf{17.79} &
            &
            &
            &
            C3:\textbf{19.55} &
            22.30 \\
        SynCoBERT~\cite{wang2021syncobert} &
            TE &
            \Block{}{Code + Doc +\\ AST Seq} &
            \Block{}{MLM + IT + TEP +\\ MCL} &
            Multi &
            D1:64.5 &
            B1:\textbf{97.4} &
            &
            P1:88.24 &
            \Block{}{C3:74.0\\ A1:38.1} &
            &
            C5:60.85 &
            &
            &
            &
            &
            &
            \\
        SPT-Code~\cite{niu2022sptcode} &
            TF & 
            \Block{}{Code +\\ Names +\\ AST Seq} &
            \Block{}{CAP \& MASS \&\\ MNG} &
            Multi &
            &
            &
            &
            &
            C3:71.5 &
            &
            C5:62.18 &
            14.2 &
            S1:19.09 &
            &
            &
            \Block{}{C3:15.0\\ T2:49.1\\ M1:36.1} &
            \\
        UniXcoder~\cite{guo2022unixcoder} &
            TE & 
            \Block{}{AST Seq +\\ Doc} &
            \Block{}{MLM / ULM / Seq2Seq\\ MLM / MCL / CMG} &
            Multi &
            &
            B1:95.2 &
            &
            P1:\textbf{90.52} &
            \Block{}{C3:\textbf{74.4}\\ A1:\textbf{41.3}} &
            C4:\textbf{70.1} &
            &
            &
            &
            &
            &
            C3:19.30 &
            \textbf{22.60} \\
        \bottomrule
    \end{NiceTabular}%
    }
\end{table*}

\begin{table*}[t!]
    \centering
    \caption{Categorization and description of the pre-training tasks mentioned in Table~\ref{table:codeptms}.}
    \label{table:pretrainingtasks}
    \resizebox{\textwidth}{!}{
    \begin{tabular}{l|l|l|ll}
        \toprule
            \multicolumn{2}{c|}{\textbf{Type}} & \textbf{O.} & \textbf{Task} & \textbf{Full Name and Description} \\
        \midrule
            \multicolumn{2}{c|}{\multirow{10}{*}{C/NLA}}  & \multirow{8}{*}{G.} & ULM ~\cite{svyatkovskiy2020gpt-c}       
            & Unidirectional LM: 
            conditional on words that have already appeared, maximizes the conditional probability of all next words. \\ \cline{4-5}
            \multicolumn{2}{c|}{}              &       & FNP~\cite{qi2021prophetnet-code}       
            & Future N-gram Prediction: 
            conditional on words that have appeared, maximizes the conditional probability of all next $N$ ($N>1$) words.\\ \cline{4-5}
            \multicolumn{2}{c|}{}               &      & BiLM  ~\cite{karampatsis2020scelmo}     
            & Bidirectional LM: 
            apply ULM to the input and its reversion to maximize the conditional probability of each word in both directions.\\ \cline{4-5}           
            \multicolumn{2}{c|}{} & & MLM ~\cite{de2021javabert}  
            & Masked Language Model: 
            predicts a certain percentage of tokens that have been randomly masked in the input. (Basic version of MLM) \\ \cline{4-5}
            \multicolumn{2}{c|}{}             &        & WWM ~\cite{buratti2020c-bert}
            & Whole Word Masking: 
            a variant of basic MLM, if parts of a word is masked, ensure all subwords/tokens in it be masked.  \\ \cline{4-5}
            \multicolumn{2}{c|}{}            &         & MASS ~\cite{niu2022sptcode}      
            & MAsked Seq2Seq: 
            predicts 50\% of the content that is randomly masked consecutively in the sentence in the encoder-decoder architecture.\\ \cline{4-5}
            \multicolumn{2}{c|}{}             &        & SMLM ~\cite{mastropaolo2021t5-learning}      
            & Seq2Seq MLM: 
            sequentially predicts a set of token spans randomly masked in the input in the encoder-decoder framework. \\ \cline{4-5}
            \multicolumn{2}{c|}{}            &      & DAE  ~\cite{ahmad2021plbart}       
            & Denoising Auto-Encoding: 
			recovers the original input from the one tampered by masking, deleting, and replacing tokens, etc.\\ \cline{3-5}
            \multicolumn{2}{c|}{} & \multirow{2}{*}{C.} & NSP ~\cite{kanade2020cubert}     
            & Next Sentence Prediction: determines whether the two given sentences or logical lines of code appear consecutively in real world. \\ \cline{4-5}
            \multicolumn{2}{c|}{}            &         & RTD ~\cite{feng2020codebert}   
            & Replaced Token Detection: identifies whether a token in the input is a fake one that is produced by a small generator network. \\ \hline
            
            \multicolumn{2}{c|}{\multirow{4}{*}{CA}} &\multirow{2}{*}{G.} & IMLM ~\cite{liu2020cuglm}      
            & Identifier MLM: 
			predicts a certain percentage of identifiers that randomly masked in the code text (an adaption of basic MLM to code). \\ \cline{4-5}
            \multicolumn{2}{c|}{}                &     & SIMLM  ~\cite{wang2021codet5}  
            & Seq2Seq IMLM: an adaptation of Seq2Seq MLM to source code that masks only a certain percentage of the identifiers in the code text.
			\\ \cline{3-5}
            \multicolumn{2}{c|}{}                &  \multirow{2}{*}{C.}   & IT/IP  ~\cite{wang2021codet5}  
            & Identifier Tagging/Predicting: determines whether the input token at each position is an identifier or not via binary classification.\\ \cline{4-5}
            \multicolumn{2}{c|}{}             &        & CCL ~\cite{peng2021oscar}       
            & Code Contrastive Learning: minimizes/maximizes the distances between the representations of similar/dissimilar code snippets.\\ \hline
            \multicolumn{2}{c|}{\multirow{2}{*}{SA}} & \multirow{2}{*}{C.} & EP/TEP ~\cite{guo2021graphcodebert}     
            & Edge Prediction: 
			predicts the edges that are masked by randomly selecting source and target nodes in a DFG or AST. \\ \cline{4-5}
            \multicolumn{2}{c|}{}                   &  & NOP ~\cite{jiang2021treebert}      
            & Node Order Prediction: 
			determines if a change occurs in an AST where the order of some randomly selected nodes are changed. \\ \hline
            
            \multirow{10}{*}{CMA} &\multirow{2}{*}{CN} & \multirow{2}{*}{G.} & BDG/CMG ~\cite{wang2021codet5} 
            &Bimodal Dual Generation/Cross Modal Generation: generates a Natural Language/Code if Code/Natural Language is given. \\ \cline{4-5}
                                   &           &            & MNG ~\cite{niu2022sptcode}   
            &Method Name Generation: produces a name for the given method body by generating sub-tokens with the decoder sequentially. \\ \cline{2-5}
            &\multirow{7}{*}{CS}  & \multirow{2}{*}{G.} & CLR  ~\cite{zhang2021codedisen}
            &Cross-Language Reconstruction: generates a code snippet in one PL functionally equivalent to the given one in other PLs. \\ \cline{4-5}
                                   &          &            & TMLM ~\cite{jiang2021treebert} 
            &Tree MLM: 
			generates complete code from inputs where some terminal nodes/identifiers in ASTs/code are masked in encoder/decoder. \\ \cline{3-5}
                                   &           &     \multirow{5}{*}{C.}      & VGVAE ~\cite{zhang2021codedisen} 
            & vMF-Gaussian Variational Autoencoder: disentangles code semantics from code syntax under the supervision of a masked AST.  \\ \cline{4-5}
                                   &          &            & CAP ~\cite{niu2022sptcode} 
            & Code-AST Prediction: determines whether the given code and AST in the input correspond to each other via binary classification.\\ \cline{4-5}
                                   &           &           & NA ~\cite{guo2021graphcodebert} 
            &Node Alignment: 
            predicts the masked edges connecting randomly sampled nodes in a DFG and its corresponding code token. \\ \cline{4-5}
                                   &          &            & PD ~\cite{zhang2021codedisen} 
            & Posterior Distribution: minimizes the difference in semantics distributions of functionally equivalent code snippets in different PLs.
            \\ \cline{4-5}
                                   &           &           & ACP ~\cite{zhang2021codedisen} 
            &Attentive Code Position: predicts the node type in AST of a code token in the input through an attention mechanism. \\ \cline{2-5}
                                   &CNS       &    C.     & MCL ~\cite{wang2021syncobert} 
            &Multi-modal Contrastive Learning: an adaptation of CCL to (Code,NL)/(Code,Structure) pairs where samples are no longer code pairs. \\ 
        \bottomrule
    \end{tabular}
    }
   \vspace{-2mm}
\end{table*}

Table~\ref{table:codeptms} presents an overview of 26 of the PTMs that are either commonly used and/or developed in SE. As can be seen, these PTMs can be divided into three groups: PTM-NL, PTM-C, and CodePTM. Within each group, we order them chronologically (by the date of the preprint or the official publication). To enable the reader to better understand their similarities and differences, we categorize the PTMs of source code (i.e., PTM-Cs and CodePTMs) along the four dimensions proposed by Niu et al.~\cite{niu2022deep}\footnote{Note that three of these four dimensions are also applicable to PTM-NL.}:

(1) Architecture (\textbf{Arch.}). Existing network architectures can be divided into \textit{Long Short-Term Memory} (LSTM)~\cite{hochreiter1997lstm}, \textit{Transformer} (TF)~\cite{vaswani2017transformer}, \textit{Transformer-Encoder} (TE, the encoder-only portion of TF), and \textit{Transformer-Decoder} (TD, the decoder-only portion of TF).

(2) \textbf{Modality} refers to the type of input a PTM assumes. The possible modalities  include \textit{code}, natural language (\textit{NL}) and code \textit{structure}. How these different modalities should be combined is determined by the underlying combination strategy, which can be \textit{together} (+) or \textit{standalone} (\&)\footnote{See the supplementary file for details.}.

(3) Pre-training Tasks (\textbf{Tasks}). If more than one task is used, the tasks can be learned \textit{jointly} (+), \textit{sequentially} (\&), or \textit{alternately} (/)\footnote{See the supplementary file for details on the different ways of pre-training a model when more than one pre-training task is involved.}. The definition of each pre-training task is given in Table~\ref{table:pretrainingtasks}. Following Niu et al,~\cite{niu2022deep}, in the first column we classify these tasks into four categories according to their input modalities: (1) Code-Aware or Natural-Language-Aware (\textbf{C/NLA}) tasks, which are originated in NLP and can be applied to either NL or Code sequence to mine latent information from NL or Code; (2) Code-Aware Only (\textbf{CA}) tasks, which can only be applied to mine latent information from code text; (3) Structure-Aware Only (\textbf{SA}) tasks, which aim to learn representations of the code structure; and (4) Cross-Modal-Aware (\textbf{CMA}) tasks, which seek to acquire knowledge from multiple input modalities and are further subdivided into three categories based on which input modalities are involved, namely Code-NL (\textbf{CN}), Code-Structure (\textbf{CS}), and Code-NL-Structure (\textbf{CNS}). In the second column, we classify these tasks based on whether they are Generative (\textbf{G}, i.e., generate tokens) or Categorical (\textbf{C}, i.e., predict labels) in nature.

(4) Programming language (\textbf{PL}). We categorize code PTMs depending on whether they are pre-trained on one PL (Monolingual (\textit{Mono})) or multiple PLs (Multilingual (\textit{Multi})).

In our empirical comparison, we exclude all LSTM-based models of source code since they do not represent the state of the art, and retain all the Transformer-based models of source code shown in Table~\ref{table:codeptms} except OSCAR, because OSCAR does not target high-level PLs. This leaves us with 19 PTMs of source code in our comparison. In addition, we will present results of five models that are {\em not} pre-trained on source code. They include four PTMs-NL (RoBERTa, GPT-2, BART, and T5) and a vanilla Transformer model~\cite{vaswani2017transformer}. Comparing the results of these models and those obtained by the 19 PTMs of source code could shed some light on the gains that can be obtained on each SE task via pre-training on source code.

\subsection{Implementation Details}
\label{section:setup_implementation}

\paragraph{The 19 PTMs of source code}
According to the public availability of the artifacts, the 19 models of source code we use in our comparison can be divided into four categories:

(1) For those PTMs that have publicly available pre-trained models and tokenizers, we use them as provided. CuBERT, CodeBERT, GraphCodeBERT, DOBF, JavaBERT, CodeGPT-adapted, T5-learning, PLBART, ProphetNet-Code, CoTexT, CodeT5, SPT-Code and UniXcoder are in this category. If more than one model is provided, we choose the ``base'' version consistent with the approach in the original paper.

(2) Of the remaining PTMs, if the source code and datasets are provided, we re-train them according to the setting introduced in the original papers to get the pre-trained models and the tokenizers. TreeBERT is the only model in this category.

(3) For those that have the source code but not the datasets, we collect the required datasets ourselves in the same way as the original authors did, and re-train them according to the settings in the original papers. Only CugLM is in this category.

(4) If no source code is provided, we re-implement and pre-train according to the settings (e.g., tokenizer, hyper-parameters, and dataset) described in the original papers. They are GPT-C, C-BERT, DeepDebug and SynCoBERT\footnote{To verify the validity of the latter two types of models pre-trained by us, we perform fine-tuning on the downstream tasks corresponding to the original paper and use pair-wise $t$-tests to ensure that the difference between our results and those reported in the original papers are statistically indistinguishable. Details can be found in the supplementary materials.}.

When evaluating on a downstream SE task, each of the 19 models is fine-tuned on the training data available for that task.

\paragraph{The 5 non-PTMs}
As noted above, we also include four PTMs-NL (RoBERTa, GPT-2, BART, and T5) and a vanilla Transformer model in our comparison. For the four PTMs-NL, we use their publicly available implementations. Like the 19 PTMs of source code, these five models are being fine-tuned on task-specific data~\cite{vaswani2017transformer} before applying to each downstream task.

\subsection{Application to SE Tasks}
\label{section:setup_application}

Two aspects need to be considered while applying PTMs to SE tasks, namely, Inputs and Outputs.

\subsubsection{Inputs}

The inputs for different SE tasks are different. When applying a PTM to a SE task, the input of the task should be organized into a form needed by the PTM. The input of the SE tasks in Table~\ref{table:codeptms} belongs to three types:

(1) \textbf{Using only a code snippet as input}: Tasks such as Defect Detection and Code Translation assume input that belongs to this category. Here, we follow the input representation as defined by PTMs. For example, for TreeBERT, we parse the code into an AST and encode each path in the AST before passing it to the Transformer, as described in the original paper; and for PLBART, we add a special symbol indicating the programming language, e.g., ``[java]'', to the input sequence. 

(2) \textbf{Using only a natural language description as input}: This is used by tasks such as Code Search and Code Generation. In this case, we input the text sequence directly. But for PLBART, we follow the approach described in its paper and add a special symbol ``[en]'' to the input. 

(3) \textbf{Using a code-code pair or a code-NL pair as input}: Tasks like Clone Detection (inputs: code-code) and Code Question Answering (inputs: code-NL) belong to this type. In this case, we prepare the inputs for the two parts separately and then concatenate them to obtain the final input representation.

\subsubsection{Outputs}

The output required by a SE task may not be the same as the output produced by the PTMs. Hence, additional modules or operations may be needed in order to get the output required by SE Tasks.
The outputs that need to be provided by PTMs for different SE tasks can be divided into two types:

(1) \textbf{Output based on the input representation}: Among the SE tasks, Code Search and Code Question Answering use the input representation directly (to calculate the similarity between two sequences), while the others need a fully connected layer and a softmax layer to be added to obtain a probability distribution. PTMs with different architectures use different ways to get the representation vector for the input. For \textbf{TE-based} models, we use the vector that corresponds to the position of the classification symbol in the input (typically ``[CLS]'') as the representation vector. For \textbf{TD-based} models, we use the last time step of the output hidden state (i.e., the position of the special symbol ``[endoftext]'' in the input sequence). For \textbf{TF-based} models, it depends. Since T5-based models (i.e., T5, T5-learning, DeepDebug, CoTexT and CodeT5) formalize all tasks as text-to-text tasks, for {\em classification tasks} we map all categories to text (e.g. for a binary classification task, 0 is mapped to ``false'' and 1 to ``true''), while for retrieval tasks, we use the output hidden state of the decoder corresponding to the ``[EOS]'' symbol as the representation vector. In contrast, for BART-based models (i.e., BART, PLBART and SPT-Code), we keep the input of the decoder to be the same as the input of the encoder and use the decoder hidden state of the last timestep as the representation vector. 
For other \textbf{TF-based} models, we only use its encoder and adopt the same method as used in the \textbf{TE-based} models.

(2) \textbf{Output based on the ultimate output sequence}: For \textbf{TE-based} models, we follow Lu et al.~\cite{lu2021codexglue} to randomly initialize a Transformer Decoder of the same size as the model to form an encoder-decoder architecture. For \textbf{TD-based} models, we follow GPT-2~\cite{radford2019gpt2}: for training, we concatenate the input and output sequences using a special symbol; and for evaluation, we pass the input sequence concatenated with this special symbol into the model and use the sequence predicted by the model as the output. \textbf{TF-based} models can be applied directly to this type of tasks. The \textbf{Code Completion} task deserves special mention. Recall that it requires a model to complete the unfinished line given the previous context. However, during training, it follows the GPT-like, casual language modeling manner. This is not applicable to TE- and TF-based PTMs that adopt the encoder-decoder architecture for this task. Therefore, when training TE- and TF-based PTMs, we randomly extract the first 15\%-85\% of the entire sequence as input (since the input context in the test data is ensured to be at least 15\% of the whole length~\cite{lu2021codexglue}) of the encoder, and the rest is used as the input of the decoder.

\subsection{Other Settings and Data Availability}
\label{section:setup_others}

For other settings, e.g., the hyperparameters and the optimizer, we adopt those used in the provided source code or mentioned in the original paper. If neither of the above is available, we perform parameter tuning ourselves to maximize model performance on held-out development data\footnote{Refer to the supplementary materials for details.}.

\section{Evaluation of PTMs: The Status Quo}
\label{section:setup_codeptms_discussion}

The current state of research on applying PTMs, including PTMs-NL, PTMs-C, and CodePTMs, to SE tasks is somewhat unsatisfactory. To understand this status quo, we show in Table~\ref{table:codeptms} the ID of each dataset  that each PTM is evaluated on (see Table~\ref{table:se_tasks} for an explanation of the dataset IDs) and the corresponding results as reported in the original papers. To avoid overloading the reader with information, we (1) omit the dataset ID when the SE task has only one dataset; (2) report results in terms of percentage using the first evaluation metric (see Table~\ref{table:se_tasks}); and (3) average results over all data subsets when a dataset is composed of multiple subsets\footnote{The datasets that comprise multiple subsets are: \texttt{C3} (which includes subsets corresponding to six languages), \texttt{C5} (which contains two subsets corresponding to bidirectional translation between Java and C\#), \texttt{B2} (which consists of two subsets of different length distributions), \texttt{A3} (which includes two subsets corresponding to raw and abstracted source code~\cite{watson2020assert}) and \texttt{C7} (which consists of three subsets of different sizes).}.

Below we discuss the status quo based on the results shown in Table~\ref{table:codeptms}, focusing our discussion on PTMs of source code given that they are the focus of this paper.

\paragraph{Code Understanding Tasks}
Among the code understanding tasks, only one PTM of source code is evaluated for Exception Type and Code Question Answering, so we have no idea of the performance of the other models on these tasks. Although three CodePTMs are evaluated on Code-to-Code Retrieval, they all used different datasets, thus making direct comparisons impossible. Consequently, the only tasks for which we can compare PTMs of source code are Defect Detection, Clone Detection and Code Search. For Defect Detection, most of the models are evaluated on Devign with CoTexT achieving the best results. For Clone Detection, most models are evaluated on BigCloneBench with SynCoBERT achieving the best results. For Code-to-Code Retrieval (POJ-104) and Code Search (CodeSearchNet and AdvTest), UniXcoder is the state-of-the-art CodePTM.

\paragraph{Code Generation Tasks}
Code generation tasks are more popularly used to evaluate PTMs of source code. However, we cannot make valid comparisons on three of the seven generation tasks shown in Table~\ref{table:codeptms}: four PTMs of source code were evaluated on Code Completion, but they each used different datasets. Only T5-learning was evaluated on Mutant Generation and Assert Generation. Of the four comparable tasks and datasets, CodeT5 achieved the best results in three of them: Code Translation (CodeTrans), Bug Fixing, and Code Summarization (CodeSearchNet). The remaining task, Code Generation, is bested by UniXcoder.

\paragraph{Overall}
Since different PTMs of source code are evaluated on different downstream tasks and dataset, it is impossible to compare them directly based only on the results reported in existing paper. Consequently, we cannot draw conclusions that are more broadly applicable, the conclusions we draw are not reliable, and we could not know the best performing PTMs of source code on many of these SE tasks. Therefore, achieving a fair and systematic comparison of these PTMs is the main motivation for the work in this paper.

\section{Evaluation Results}
\label{section:results}

In this section, we present our evaluation results.

For each SE task, we repeat the fine-tuning and testing experiments on each  model three times using three random seeds (i.e., 24, 42 and 81) and report the average results in Tables~\ref{table:results_und}, \ref{table:results_gen_ct_ag}, \ref{table:results_gen_bf_cc_mg}, and \ref{table:results_gen_nl}. Specifically, for each task and each evaluation metric of that task, we show under the ``Cur'' and ``New'' columns the current results reported by existing work and the results we obtained through our experiments, respectively. For each ``New'' column, the best and second best results obtained by each type of pre-trained models (i.e., PTM-NL, PTM-C, and CodePTM) on the corresponding SE task are marked in \textbf{bold} and \underline{underline} respectively\footnote{For each SE task, we apply the Approximate Randomization Test ~\cite{edgington1969approximate} to every pair of results to determine whether their difference is statistically significant. These results are presented in the supplementary materials.}. Note that the first row of each table show the results of vanilla Transformer.

\begin{table}[t!]
    \centering
    \caption{Current SOTAs and new SOTAs.}
    \label{table:sota_compare}
    \resizebox{\linewidth}{!}{%
    \begin{NiceTabular}{lllllr}
    \CodeBefore
            \rowcolors{4}{}{gray!15}
            \Body
        \toprule
            &
            \Block{1-2}{\textbf{Current SOTA}} &
            &
            \Block{1-2}{\textbf{New SOTA}} & 
            &
            \Block[c]{}{\textbf{$\Delta$}} \\
        \cline{2-5}
            &
            \textbf{Model} &
            \textbf{Value (\%)} &
            \textbf{Model} &
            \textbf{Value (\%)} &
            \textbf{(pts)} \\
        \midrule
            DD &
            CoTexT &
            65.99 &
            \textbf{SynCoBERT} &
            66.25 &
            0.26 \\
        CD &
            SynCoBERT &
            97.4 &
            SynCoBERT &
            97.55 &
            0.15 \\
        ET &
            CuBERT &
            79.12 &
            \textbf{CodeT5} &
            85.00 &
            5.88 \\
        CR &
            UniXcoder &
            90.52 &
            UniXcoder &
            90.55 &
            0.03 \\
        CS &
            UniXcoder &
            41.3 &
            UniXcoder &
            41.57 &
            0.27 \\
        QA &
            UniXcoder &
            70.1 &
            UniXcoder &
            70.3 &
            0.2 \\
        CT &
            CodeT5 &
            66.4 &
            \textbf{PLBART} &
            67.6 &
            1.2 \\
        BF &
            CodeT5 &
            17.79 &
            CodeT5 &
            17.98 &
            0.19 \\
        CC &
            CodeGPT-adapted &
            42.37 &
            CodeGPT-adapted &
            43.80 &
            1.43 \\
        MG &
            T5-learning &
            28 &
            \textbf{CodeT5} &
            34.83  &
            6.83 \\
        AG &
            T5-learning &
            40.5 &
            \textbf{PLBART} &
            49.94 &
            9.44 \\
        SM &
            CodeT5 &
            19.55 &
            CodeT5 &
            19.71 &
            0.16 \\
        CG &
            UniXcoder &
            22.60 &
            \textbf{CodeT5} &
            23.43 &
            0.83 \\
        \bottomrule
    \end{NiceTabular}
    }%
\end{table}
\begin{table*}[htbp]
    \centering
    \caption{Experimental results on code understanding tasks.}
    \label{table:results_und}
    \resizebox{0.68\linewidth}{!}{%
        \begin{NiceTabular}{lrrrrrrrrrrrr}
        \CodeBefore
            \rowcolors{4}{gray!15}{}
        \Body
            \toprule
                \Block{3-1}{\textbf{Model}} &
                \Block[c]{1-2}{\textbf{DD}} &
                &
                \Block[c]{1-2}{\textbf{CD}} &
                &
                \Block[c]{1-2}{\textbf{ET}} &
                &
                \Block[c]{1-2}{\textbf{CR}} &
                &
                \Block[c]{1-2}{\textbf{CS}} &
                &
                \Block[c]{1-2}{\textbf{QA}} &
                \\
            \cline{2-13}
                &
                \Block[c]{1-2}{\textbf{Acc}} &
                &
                \Block[c]{1-2}{\textbf{F1}} &
                &
                \Block[c]{1-2}{\textbf{Acc}} &
                &
                \Block[c]{1-2}{\textbf{MAP}} &
                &
                \Block[c]{1-2}{\textbf{MRR}} &
                &
                \Block[c]{1-2}{\textbf{MRR}} &
                \\
            \cline{2-13}
                &
                \textbf{Cur} &
                \textbf{New} &
                \textbf{Cur} &
                \textbf{New} &
                \textbf{Cur} &
                \textbf{New} &
                \textbf{Cur} &
                \textbf{New} &
                \textbf{Cur} &
                \textbf{New} &
                \textbf{Cur} &
                \textbf{New} \\
            \midrule
                Transformer &
                &
                64.40 &
                &
                89.27 &
                &
                48.98 &
                &
                64.27 &
                &
                3.12 &
                &
                52.89 \\
            \hline
                RoBERTa &
                61.05 &
                \textbf{64.47} &
                94.9 &
                \underline{95.35} &
                - &
                \textbf{76.94} &
                76.67 &
                \textbf{80.20} &
                18.33 &
                \textbf{18.82} &
                60.3 &
                \textbf{60.28} \\
            GPT-2 &
                - &
                63.22 &
                - &
                \textbf{96.22} &
                - &
                \underline{75.54} &
                - &
                53.30 &
                - &
                16.38 &
                - &
                \underline{58.06} \\
            BART &
                - &
                \underline{63.81} &
                - &
                95.11 &
                - &
                73.68 &
                - &
                \underline{79.63} &
                - &
                16.65 &
                - &
                55.57 \\
            T5 &
                61.93 &
                61.87 &
                - &
                94.86 &
                - &
                74.75 &
                - &
                69.16 &
                - &
                \underline{16.97} &
                - &
                45.63 \\
            \hline
                CuBERT &
                - &
                64.25 &
                - &
                94.78 &
                79.12 &
                \textbf{79.90} &
                - &
                \underline{76.87} &
                - &
                22.26 &
                - &
                54.33 \\
            GPT-C &
                - &
                63.77 &
                - &
                95.46 &
                - &
                \underline{78.26} &
                - &
                55.23 &
                - &
                24.39 &
                - &
                50.32 \\
            C-BERT &
                57.4 &
                64.05 &
                - &
                95.00 &- 
                - &
                74.57 &
                - &
                72.91 &
                - &
                25.34 &
                - &
                \underline{54.81} \\
            JavaBERT &
                - &
                \underline{64.50} &
                - &
                \underline{96.57} &
                - &
                67.66 &
                - &
                \textbf{77.44} &
                - &
                25.02 &
                - &
                54.04 \\
            CodeGPT-adapted &
                - &
                \textbf{65.64} &
                - &
                \textbf{96.65} &
                - &
                76.71 &
                - &
                72.63 &
                - &
                \underline{25.97} &
                - &
                54.24 \\
            DeepDebug &
                - &
                64.18 &
                - &
                95.90 &
                - &
                73.50 &
                - &
                73.51 &
                - &
                \textbf{30.58} &
                - &
                \textbf{57.39} \\
            \hline
                CodeBERT &
                - &
                65.02 &
                - &
                96.77 &
                - &
                81.25 &
                82.67 &
                85.61 &
                - &
                38.21 &
                65.7 &
                65.90 \\
            GraphCodeBERT &
                - &
                \underline{65.92} &
                97.1 &
                97.11 &
                - &
                83.26 &
                85.16 &
                87.73 &
                - &
                38.76 &
                68.4 &
                68.55 \\
            CugLM &
                - &
                64.19 &
                - &
                96.44 &
                - &
                79.01 &
                - &
                83.32 &
                - &
                36.20 &
                - &
                61.44 \\
            DOBF &
                - &
                63.86 &
                95.9 &
                96.84 &
                - &
                79.04 &
                - &
                87.31 &
                38.3 &
                38.56 &
                - &
                61.31 \\
            T5-learning &
                - &
                63.60 &
                - &
                96.38 &
                - &
                69.85 &
                - &
                80.82 &
                - &
                37.98 &
                - &
                60.21 \\
            PLBART &
                63.18 &
                64.21 &
                97.2 &
                97.01 &
                - &
                77.93 &
                - &
                85.02 &
                - &
                38.70 &
                65.0 &
                65.01 \\
            ProphetNet-Code &
                - &
                63.57 &
                - &
                96.05 &
                - &
                79.37 &
                - &
                79.82 &
                - &
                37.64 &
                - &
                63.73 \\
            CoTexT &
                65.99 &
                65.68 &
                - &
                95.96 &
                - &
                77.21 &
                - &
                86.65 &
                - &
                38.13 &
                - &
                68.70 \\
            TreeBERT &
                - &
                65.76 &
                - &
                96.51 &
                - &
                78.08 &
                - &
                85.54 &
                - &
                39.60 &
                - &
                64.98 \\
            CodeT5 &
                65.78 &
                65.82 &
                97.2 &
                \underline{97.18} &
                - &
                \textbf{85.00} &
                - &
                87.53 &
                - &
                \underline{40.03} &
                67.8 &
                67.91 \\
            SynCoBERT &
                64.5 &
                \textbf{66.25} &
                97.4 &
                \textbf{97.55} &
                - &
                82.70 &
                88.24 &
                \underline{88.52} &
                38.1 &
                39.99 &
                - &
                \underline{69.19} \\
            SPT-Code &
                - &
                64.88 &
                - &
                96.40 &
                - &
                77.11 &
                - &
                86.54 &
                - &
                37.05 &
                - &
                64.55 \\
            UniXcoder &
                - &
                65.64 &
                95.2 &
                96.32 &
                - &
                \underline{83.47} &
                90.52 &
                \textbf{90.55} &
                41.3 &
                \textbf{41.57} &
                70.1 &
                \textbf{70.30} \\
            \bottomrule
        \end{NiceTabular}%
     }
\end{table*}

\section{Discussion}
\label{section:results_discussion_task}

Through our experiments, we obtain the new SOTA\footnote{For generation tasks (e.g., Code Translation, Assert Generation, Bug Fixing, Code Completion, etc.) adopting multiple metrics (e.g., EM, BLEU, CodeBLEU, etc.), we follow the existing papers and use EM (Exact Match) as the metric to determine the SOTA model. The reason is that EM is the most rigorous metric for evaluating the generation performance. Besides, if there are multiple sub-datasets for evaluating one task, the average performance over all sub-datasets is used to determine the SOTA model.} on each task. In Table~\ref{table:sota_compare} we show for each task the current SOTA model (derived from existing work), the new SOTA model (derived from our experiments), as well as their corresponding performances\footnote{If multiple sub-datasets are used for evaluation, the average performance is reported here.}. Moreover, we boldface the new SOTA model if it is different from the current SOTA model. Finally, we show the absolute performance difference between the new SOTA model and the current SOTA model under the ``$\Delta$'' column.

First, the SOTA models of all 13 SE tasks belong to the type of CodePTM, which covers models specifically designed to capture the unique features of Source Code, except for the Code Completion task whose SOTA model, CodeGPT-adapted, is of type PTM-C, which covers models designed for Natural Language but pre-trained on Source Code. 

Second, while many PTMs have been proposed, only five of them have managed to achieve SOTA performance on at least one SE task. They are CodeT5 (SOTA on 5 tasks), UniXcoder (SOTA on 3 tasks), PLBART (SOTA on 2 tasks), SynCoBERT (SOTA on 2 tasks), and CodeGPT-adapted (SOTA on 1 task). 

Third, vanilla Transformer's performance relative to the PTMs is different for different SE tasks: (1) on Clone Detection (CD), Error Type prediction (ET), Code Search (CS), Code Translation, Assert Generation, and Code Summarization, vanilla Transformer is surpassed in performance by all types of PTMs (i.e., PTM-NL, PTM-C, and CodePTM); (2) on Code Completion and Mutant Generation, vanilla Transformer is beaten by all PTMs-C and CodePTMs but it outperforms two PTMs-NL, BART and T5; (3) on Code-to-Code Retrieval (CR) and Code Question Answering (QA), vanilla Transformer not only surpasses a PTM-NL (GPT-2 on CR and T5 on QA), but also beats one PTM-C (GPT-C for both tasks); and (4) on Defect Detection (DD) and Bug Fixing, vanilla Transformer even outperforms CodePTMs in addition to PTMs-NL and PTMs-C, beating CugLM, DOBF, T5-learning, PLBART, and ProphetNet-Code on DD, and CugLM on Bug Fixing. 

In the following, we discuss in detail the observations obtained from the current and the new results on each task.

\subsection{Defect Detection}

SynCoBERT defeats CoTexT and becomes the new SOTA PTM for this task, and Accuracy improves by 0.26. 

\subsubsection{Architecture} While the Top-2 models  on this classification task, SynCoBERT and GraphCodeBERT, are both TE-based, \textbf{there is not enough empirical evidence for us to conclude that TE is a better architecture for this task than TD or TF}, for several reasons. First, to draw this conclusion, we need to compare the results of two models that differ only w.r.t.\ architecture, but there do not exist two PTMs on our list that differ only w.r.t.\ architecture. Second, TF-based CoTexT, which uses MLM as the only pre-traning task, outperforms TE-based UniXcoder, which uses three more complex pre-training tasks, ULM, MCL and CMG. Finally, TF-based DeepDebug achieves better results than TE-based C-BERT when using only code as input and MLM as its only pre-training task. 

\subsubsection{Modality} \textbf{Both code structure and NL are shown to have a positive effect on the performance of the models on this task, but the way they are being used also matters.} As an example, TF-based TreeBERT outperforms some of the TF-based models that use code and NL (e.g, DOBF, T5-learning, PLBART) significantly owning to its use of ASTs. As another example, TF-based CoTexT outperforms TF-based T5-learning considerably: CoText concatenates Code and the corresponding Doc as one single input, whereas T5-learning treats the features derived from these two modalities as separate data instances. This suggests that how the information derived from these modalities is used has an impact on performance.

\subsubsection{Pre-training Tasks} First, the results in the \textit{New} column of this task in Table~\ref{table:results_und} reveal that \textbf{the most influential pre-training tasks are cross-modal-aware classification tasks as they are being used by the Top-5 models}. These tasks include \textit{TEP/EP} (used by SynCoBERT and GraphCodeBERT), \textit{MCL} (used by SynCoBERT and UniXcoder), and \textit{NA} (adopted by GraphCodeBERT). This observation is different from the conclusion derived from the \textit{Cur} column, where Seq2Seq MLM (the only pre-training task used by the old SOTA model, CoTexT) seems to have the greatest impact on defect detection. 

\subsection{Clone Detection}

The new results on this task do not change significantly from the current ones, except that PLBART, which is currently tied for second place, has slipped to fourth place. The drop in PLBART's rank seems to suggest that using multiple pre-training tasks is better than using a single pre-training task on this task: while the Top-2 models, SynCoBERT and CodeT5, employ four distinct pre-training tasks, PLBART uses DAE as the only pre-training task. Besides, the new results also enable us to see the performance of TD-based models on this task; in particular, the best TD-based PTM, CodeGPT-adapted, ranks 7th.

\subsection{Exception Type}

This task is the only multi-label classification task among our 13 SE evaluation tasks. Currently, only one model (i.e., CuBERT) has been applied to this task, which prevents us from drawing any conclusions about the relative performance of different types of models on a multi-label classification task like this. Fortunately, our results enable us to draw several new conclusions:

\subsubsection{Architecture} Most notably, according to the new results, \textbf{the SOTA performance on this task is not achieved by a TE-based model}. Instead, TF-based CodeT5, which turns the task into a text-to-text form, achieves the best results. The best TE-based model (UniXcoder) and the best TD-based model (GPT-C) rank second and tenth respectively, and their accuracies are 1.53 and 6.74 points lower than that of CodeT5. Recall that in Section~\ref{section:setup_application}, we mentioned that as a T5-based model, CodeT5, when applied to a classification task, maps each label to a unique text string. Specifically for Exception Type, it does not predict the index of each exception, but rather the text string of that exception. In this way, CodeT5 turns this classification task into one of generating NL, which is exactly what CodeT5 is good at. In contrast, for TE-based models (e.g., SynCoBERT, UniXcoder, GraphCodeBERT), most of the tasks they use in pre-training are binary classification tasks (e.g., MCL, TEP/EP, NA), so they may lack the knowledge needed for multi-label classification.

\subsubsection{Modality} The impact of each modality on this task becomes clear as well. All of the Top-3 models (i.e., CodeT5, UniXcoder, and GraphCodeBERT) use NL as one of the input modalities, while both code and code structure were only used by two of them (CodeT5 and UniXcoder). This seems to suggest that \textbf{NL has a better positive impact on this task than the other two modalities}. 

\subsubsection{Pre-training Tasks}
\textbf{Both the classification pre-training task NSP and the generative pre-training task FNP seem to have positive impacts on this task}. To exemplify, while CuBERT and C-BERT are both TE-based models that use code as the only modality and differ only in their pre-training tasks (CuBERT uses both MLM and NSP whereas C-BERT uses only MLM), CuBERT outperforms C-BERT by as many as 5 percent points in accuracy. As another example,
while ProphetNet-Code and PLBART are both TF-based models that use code and NL as input modalities and differ only in terms of their pre-training tasks (ProphetNet-Code uses FNP whereas PLBART uses DAE), ProphetNet-Code surpasses PLBART in performance.

\subsection{Code-to-Code Retrieval}

Currently, the relative advantages and disadvantages of different model architectures are not available since only four TE-based models are evaluated on this task. However, with the new results, the conclusion that \textbf{TE-based models have more advantages over the other architectures on this task} can be verified, since the Top-3 models of this task are all TE-based (i.e., UniXcoder, SynCoBERT, and GraphCodeBERT). Besides, the performance of the TF- and TD-based models is also measurable. Specifically, the best performing TF-based model (CodeT5) and TD-based one (CodeGPT-adapted) ranks 4th and 20th, respectively. 

\subsection{Code Search}

\subsubsection{Architecture} Although the SOTA model on this task is still UniXcoder (TE-based), the rank of CodeT5 (TF-based) improved from third to second in the new results, and the third position is taken by SynCoBERT (TE-based). TreeBERT (TF-based) ranks fourth, GraphCodeBERT (TE-based) ranks fifth, and PLBART (TF-based) ranks sixth. These results seem to suggest that \textbf{TE-based and TF-based models perform comparably on this task}, as they alternate in the Top-6. In addition, the performance of TD-based models on this task is now measurable: the best TD-based PTM (CodeGPT-adapted) ranks 15th.

\begin{table*}[t!]
    \centering
    \caption{Experimental results on code translation and assert generation.}
    \label{table:results_gen_ct_ag}
    \resizebox{\linewidth}{!}{%
        \begin{NiceTabular}{lrrrrrrrrrrrr|rrrrrrrr}
        \CodeBefore
            \rowcolors{5}{}{gray!15}
        \Body
            \toprule
                \Block{4-1}{\textbf{Model}} &
                \Block{1-12}{\textbf{Code Translation}} &
                &
                &
                &
                &
                &
                &
                &
                &
                &
                &
                &
                \Block{1-8}{\textbf{Assert Generation}} &
                &
                &
                &
                &
                &
                &
                \\
            \cline{2-21}
                &
                \Block{1-6}{\textbf{Java-\textgreater{}C\#}} &
                &
                &
                &
                &
                &
                \Block{1-6}{\textbf{C\#-\textgreater{}Java}} &
                &
                &
                &
                &
                &
                \Block{1-4}{\textbf{abs}} &
                &
                &
                &
                \Block{1-4}{\textbf{raw}} &
                &
                &
                \\
            \cline{2-21}
                &
                \Block[c]{1-2}{\textbf{EM}} &
                &
                \Block[c]{1-2}{\textbf{BLEU}} &
                &
                \Block[c]{1-2}{\textbf{CodeBLEU}} &
                &
                \Block[c]{1-2}{\textbf{EM}} &
                &
                \Block[c]{1-2}{\textbf{BLEU}} &
                &
                \Block[c]{1-2}{\textbf{CodeBLEU}} &
                &
                \Block[c]{1-2}{\textbf{EM}} &
                &
                \Block[c]{1-2}{\textbf{BLEU}} &
                &
                \Block[c]{1-2}{\textbf{EM}} &
                &
                \Block[c]{1-2}{\textbf{BLEU}} &
                \\
            \cline{2-21}
                &
                \textbf{Cur} &
                \textbf{New} &
                \textbf{Cur} &
                \textbf{New} &
                \textbf{Cur} &
                \textbf{New} &
                \textbf{Cur} &
                \textbf{New} &
                \textbf{Cur} &
                \textbf{New} &
                \textbf{Cur} &
                \textbf{New} &
                \textbf{Cur} &
                \textbf{New} &
                \textbf{Cur} &
                \textbf{New} &
                \textbf{Cur} &
                \textbf{New} &
                \textbf{Cur} &
                \textbf{New} \\
            \midrule
                Transformer &
                33.0 &
                40.8 &
                55.84 &
                60.22 &
                63.74 &
                67.10 &
                37.9 &
                43.9 &
                50.47 &
                54.86 &
                61.59 &
                61.84 &
                - &
                28.65 &
                - &
                63.24 &
                - &
                35.35 &
                - &
                67.62 \\
            \hline
                RoBERTa &
                - &
                \textbf{58.9} &
                - &
                \textbf{79.70} &
                - &
                \textbf{83.77} &
                - &
                \textbf{59.5} &
                - &
                \underline{73.14} &
                - &
                \textbf{78.63} &
                - &
                \textbf{36.12} &
                - &
                \textbf{69.11} &
                - &
                \textbf{50.68} &
                - &
                \textbf{76.23} \\
            GPT-2 &
                - &
                \underline{54.4} &
                - &
                65.39 &
                - &
                \underline{76.92} &
                - &
                \underline{56.0} &
                - &
                69.41 &
                - &
                71.24 &
                - &
                \underline{35.02} &
                - &
                \underline{66.89} &
                - &
                \underline{48.23} &
                - &
                74.15 \\
            BART &
                - &
                49.5 &
                - &
                67.91 &
                - &
                74.25 &
                - &
                55.2 &
                - &
                72.32 &
                - &
                70.80 &
                - &
                33.82 &
                - &
                65.11 &
                - &
                48.14 &
                - &
                \underline{74.27} \\
            T5 &
                - &
                45.3 &
                - &
                \underline{69.23} &
                - &
                76.05 &
                - &
                53.2 &
                - &
                \textbf{73.84} &
                - &
                \underline{71.52} &
                - &
                33.10 &
                - &
                64.95 &
                - &
                48.04 &
                - &
                74.13 \\
            \hline
                CuBERT &
                - &
                55.6 &
                - &
                75.15 &
                - &
                79.30 &
                - &
                55.6 &
                - &
                70.07 &
                - &
                75.31 &
                - &
                37.14 &
                - &
                68.34 &
                - &
                50.60 &
                - &
                74.31 \\
            GPT-C &
                - &
                60.9 &
                - &
                77.91 &
                - &
                82.48 &
                - &
                59.6 &
                - &
                \underline{72.94} &
                - &
                78.18 &
                - &
                37.16 &
                - &
                66.39 &
                - &
                51.64 &
                - &
                76.12 \\
            C-BERT &
                - &
                55.4 &
                - &
                74.65 &
                - &
                81.63 &
                - &
                56.4 &
                - &
                70.28 &
                - &
                75.73 &
                - &
                36.64 &
                - &
                65.51 &
                - &
                50.70 &
                - &
                72.11 \\
            JavaBERT &
                - &
                \underline{61.1} &
                - &
                \underline{80.74} &
                - &
                80.45 &
                - &
                58.3 &
                - &
                70.10 &
                - &
                77.14 &
                - &
                38.23 &
                - &
                \textbf{71.22} &
                - &
                52.66 &
                - &
                \textbf{78.51} \\
            CodeGPT-adapted &
                - &
                \textbf{62.0} &
                - &
                80.21 &
                - &
                \textbf{85.47} &
                - &
                \underline{60.3} &
                - &
                72.93 &
                - &
                \underline{79.02} &
                - &
                \underline{39.42} &
                - &
                69.61 &
                - &
                \underline{52.85} &
                - &
                76.12 \\
            DeepDebug &
                - &
                59.5 &
                - &
                \textbf{81.46} &
                - &
                \underline{83.95} &
                - &
                \textbf{63.8} &
                - &
                \textbf{75.10} &
                - &
                \textbf{82.43} &
                - &
                \textbf{39.89} &
                - &
                \underline{71.10} &
                - &
                \textbf{56.65} &
                - &
                \underline{76.84} \\
            \hline
                CodeBERT &
                59.0 &
                61.2 &
                79.92 &
                81.16 &
                85.10 &
                85.29 &
                58.0 &
                60.1 &
                72.14 &
                73.73 &
                79.41 &
                80.11 &
                - &
                38.40 &
                - &
                70.65 &
                - &
                53.23 &
                - &
                77.54 \\
            GraphCodeBERT &
                59.4 &
                62.6 &
                80.58 &
                81.24 &
                - &
                85.34 &
                58.8 &
                61.5 &
                72.64 &
                73.67 &
                - &
                80.63 &
                - &
                38.98 &
                - &
                70.87 &
                - &
                53.71 &
                - &
                77.69 \\
            CugLM &
                - &
                60.8 &
                - &
                78.34 &
                - &
                83.65 &
                - &
                61.6 &
                - &
                73.95 &
                - &
                78.06 &
                - &
                39.36 &
                - &
                73.20 &
                - &
                52.85 &
                - &
                77.46 \\
            DOBF &
                - &
                64.8 &
                - &
                80.27 &
                - &
                82.77 &
                - &
                64.6 &
                - &
                75.44 &
                - &
                80.53 &
                - &
                40.01 &
                - &
                72.39 &
                - &
                54.71 &
                - &
                78.40 \\
            T5-learning &
                - &
                62.9 &
                - &
                78.19 &
                - &
                81.13 &
                - &
                64.8 &
                - &
                75.64 &
                - &
                81.09 &
                34 &
                40.95 &
                - &
                72.70 &
                47 &
                56.85 &
                - &
                77.09 \\
            PLBART &
                64.6 &
                \textbf{67.8} &
                83.02 &
                \underline{84.75} &
                87.92 &
                \textbf{88.16} &
                65.0 &
                \textbf{67.4} &
                78.35 &
                \textbf{79.75} &
                85.27 &
                \underline{85.05} &
                - &
                \textbf{42.44} &
                - &
                \underline{74.21} &
                - &
                \textbf{57.43} &
                - &
                79.51 \\
            ProphetNet-Code &
                - &
                62.5 &
                - &
                80.38 &
                - &
                81.64 &
                - &
                64.5 &
                - &
                75.68 &
                - &
                81.04 &
                - &
                37.38 &
                - &
                68.45 &
                - &
                56.26 &
                - &
                77.64 \\
            CoTexT &
                - &
                65.7 &
                - &
                83.35 &
                - &
                85.63 &
                - &
                65.4 &
                - &
                77.98 &
                - &
                82.31 &
                - &
                38.19 &
                - &
                71.51 &
                - &
                56.04 &
                - &
                78.55 \\
            TreeBERT &
                - &
                62.1 &
                - &
                81.72 &
                - &
                84.34 &
                - &
                64.2 &
                - &
                76.33 &
                - &
                81.17 &
                - &
                42.32 &
                - &
                73.95 &
                - &
                \underline{57.21} &
                - &
                \textbf{79.89} \\
            CodeT5 &
                65.9 &
                \underline{67.2} &
                84.03 &
                \textbf{84.97} &
                - &
                \underline{87.50} &
                66.9 &
                \underline{66.3} &
                79.87 &
                \underline{79.67} &
                - &
                83.70 &
                - &
                40.67 &
                - &
                71.77 &
                - &
                56.90 &
                - &
                \underline{79.71} \\
            SynCoBERT &
                60.4 &
                64.1 &
                80.75 &
                82.52 &
                84.85 &
                85.60 &
                61.3 &
                63.8 &
                76.52 &
                77.53 &
                82.22 &
                82.36 &
                - &
                39.10 &
                - &
                70.42 &
                - &
                54.66 &
                - &
                79.27 \\
            SPT-Code &
                64.1 &
                66.6 &
                90.34 &
                83.24 &
                - &
                85.15 &
                60.3 &
                63.9 &
                86.10 &
                78.82 &
                - &
                \textbf{85.33} &
                - &
                \underline{42.35} &
                - &
                \textbf{74.53} &
                - &
                57.09 &
                - &
                79.36 \\
            UniXcoder &
                - &
                64.5 &
                - &
                81.66 &
                - &
                85.60 &
                - &
                64.1 &
                - &
                77.37 &
                - &
                82.56 &
                - &
                39.46 &
                - &
                71.25 &
                - &
                54.94 &
                - &
                78.99 \\
            \bottomrule
        \end{NiceTabular}
    }
\end{table*}

\subsubsection{Pre-training Tasks} \textbf{The MLM pre-training task and its variants, as well as cross-modal-aware tasks demonstrate their necessity in achieving top performance on this task}. Specifically, the pre-training tasks the top-ranked models used all include \textit{MLM} (and its variants such as \textit{Seq2seq MLM}), as well as cross-modal-aware tasks (e.g., \textit{MCL}, \textit{BDG}, \textit{EP}). On one hand, \textit{MLM} and its variants enable a model to generate better input representations. On the other hand, the cross-modal-aware tasks typically allow a model to learn the alignment between different input modalities with the same semantics. These two types of pre-training tasks therefore allow a model to generate a more uniform input representation for multimodal inputs, which is exactly what a model needs to have for Code Search.

\subsubsection{Modality} \textbf{Pre-training on multiple modalities appear to benefit this task} since all of the Top-6 models are pre-trained on two or three modalities. Concretely, UniXcoder is pre-trained on NL and Structure, TreeBERT is pre-trained on Code and Structure, CodeT5 and PLBART are both pre-trained on Code and NL, while SynCoBERT and GraphCodeBERT are pre-trained on all of the three modalities. It is hard to tell which modality has the largest impact on performance, because the absence of any one of them would not prevent a model from becoming the Top-6.. 

\begin{table*}[t!]
    \centering
    \caption{Experimental results on bug fixing, code completion and mutant generation.}
    \label{table:results_gen_bf_cc_mg}
    \resizebox{\linewidth}{!}{%
        \begin{NiceTabular}{lrrrrrrrrrrrr|rrrr|rrrr}
        \CodeBefore
            \rowcolors{5}{}{gray!15}
        \Body
            \toprule
                \Block{4-1}{\textbf{Model}} &
                \Block{1-12}{\textbf{Bug Fixing}} &
                &
                &
                &
                &
                &
                &
                &
                &
                &
                &
                &
                \Block{2-4}{\textbf{Code}\\ \textbf{Completion}} &
                &
                &
                &
                \Block{2-4}{\textbf{Mutant}\\ \textbf{Generation}} &
                &
                &
                \\
            \cline{2-13}
                &
                \Block{1-6}{\textbf{small}} &
                &
                &
                &
                &
                &
                \Block{1-6}{\textbf{medium}} &
                &
                &
                &
                &
                &
                &
                &
                &
                \\
            \cline{2-21}
                &
                \Block[c]{1-2}{\textbf{EM}} &
                &
                \Block[c]{1-2}{\textbf{BLEU}} &
                &
                \Block[c]{1-2}{\textbf{CodeBLEU}} &
                &
                \Block[c]{1-2}{\textbf{EM}} &
                &
                \Block[c]{1-2}{\textbf{BLEU}} &
                &
                \Block[c]{1-2}{\textbf{CodeBLEU}} &
                &
                \Block[c]{1-2}{\textbf{EM}} &
                &
                \Block[c]{1-2}{\textbf{ES}} &
                &
                \Block[c]{1-2}{\textbf{EM}} &
                &
                \Block[c]{1-2}{\textbf{BLEU}} \\
            \cline{2-21}
                &
                \textbf{Cur} &
                \textbf{New} &
                \textbf{Cur} &
                \textbf{New} &
                \textbf{Cur} &
                \textbf{New} &
                \textbf{Cur} &
                \textbf{New} &
                \textbf{Cur} &
                \textbf{New} &
                \textbf{Cur} &
                \textbf{New} &
                \textbf{Cur} &
                \textbf{New} &
                \textbf{Cur} &
                \textbf{New} &
                \textbf{Cur} &
                \textbf{New} &
                \textbf{Cur} &
                \textbf{New} \\
            \midrule
                Transformer &
                14.7 &
                14.63 &
                77.21 &
                76.92 &
                - &
                73.88 &
                3.7 &
                8.46 &
                89.25 &
                89.23 &
                - &
                86.86 &
                - &
                37.71 &
                - &
                67.95 &
                - &
                24.50 &
                - &
                78.60 \\
            \hline
                RoBERTa &
                - &
                13.88 &
                - &
                \textbf{79.72} &
                - &
                \textbf{78.31} &
                - &
                \textbf{9.09} &
                - &
                \textbf{88.69} &
                - &
                \textbf{84.05} &
                - &
                \underline{39.01} &
                - &
                \underline{68.98} &
                - &
                \textbf{28.18} &
                - &
                \textbf{80.34} \\
            GPT-2 &
                - &
                \underline{14.97} &
                - &
                64.42 &
                - &
                68.10 &
                - &
                5.05 &
                - &
                74.11 &
                - &
                72.42 &
                41.73 &
                \textbf{41.77} &
                - &
                \textbf{70.30} &
                - &
                \underline{26.77} &
                - &
                \underline{79.45} \\
            BART &
                16.7 &
                \textbf{15.60} &
                - &
                \underline{71.34} &
                - &
                72.43 &
                6.7 &
                \underline{6.97} &
                - &
                \underline{82.37} &
                - &
                \underline{81.85} &
                - &
                30.67 &
                - &
                56.17 &
                - &
                23.21 &
                - &
                77.13 \\
            T5 &
                15.3 &
                14.34 &
                - &
                69.71 &
                - &
                \underline{73.10} &
                4.11 &
                6.49 &
                - &
                78.22 &
                - &
                78.20 &
                - &
                28.58 &
                - &
                55.24 &
                - &
                24.17 &
                - &
                79.16 \\
            \hline
                CuBERT &
                - &
                14.87 &
                - &
                74.93 &
                - &
                75.28 &
                - &
                8.92 &
                - &
                \underline{86.12} &
                - &
                83.09 &
                - &
                38.32 &
                - &
                66.97 &
                - &
                27.08 &
                - &
                78.66 \\
            GPT-C &
                - &
                13.08 &
                - &
                70.06 &
                - &
                71.83 &
                - &
                8.26 &
                - &
                85.41 &
                - &
                82.47 &
                - &
                \underline{42.82} &
                - &
                \underline{71.35} &
                - &
                27.24 &
                - &
                76.55 \\
            C-BERT &
                - &
                14.04 &
                - &
                73.19 &
                - &
                74.54 &
                - &
                9.37 &
                - &
                85.57 &
                - &
                83.87 &
                - &
                41.07 &
                - &
                67.82 &
                - &
                26.63 &
                - &
                77.43 \\
            JavaBERT &
                - &
                \underline{15.39} &
                - &
                \textbf{78.98} &
                - &
                76.02 &
                - &
                9.41 &
                - &
                85.33 &
                - &
                84.42 &
                - &
                39.16 &
                - &
                67.67 &
                - &
                \underline{28.14} &
                - &
                79.33 \\
            CodeGPT-adapted &
                - &
                13.66 &
                - &
                76.07 &
                - &
                \textbf{77.13} &
                - &
                \underline{11.00} &
                - &
                85.28 &
                - &
                \underline{84.55} &
                42.37 &
                \textbf{43.80} &
                - &
                \textbf{72.54} &
                - &
                27.64 &
                - &
                \underline{79.40} \\
            DeepDebug &
                18.7 &
                \textbf{18.13} &
                - &
                \underline{76.64} &
                - &
                \underline{76.91} &
                11.4 &
                \textbf{11.09} &
                - &
                \textbf{87.10} &
                - &
                \textbf{85.80} &
                - &
                39.68 &
                - &
                65.28 &
                - &
                \textbf{30.11} &
                - &
                \textbf{79.45} \\
            \hline
                CodeBERT &
                16.4 &
                14.66 &
                77.42 &
                78.41 &
                - &
                78.09 &
                5.2 &
                9.72 &
                90.07 &
                86.94 &
                - &
                83.88 &
                - &
                40.77 &
                - &
                68.58 &
                - &
                28.61 &
                - &
                \underline{80.59} \\
            GraphCodeBERT &
                17.3 &
                16.85 &
                80.02 &
                \textbf{79.61} &
                - &
                \textbf{79.68} &
                9.1 &
                10.14 &
                91.31 &
                87.63 &
                - &
                85.33 &
                - &
                40.26 &
                - &
                69.88 &
                - &
                29.72 &
                - &
                80.53 \\
            CugLM &
                - &
                13.78 &
                - &
                75.36 &
                - &
                75.20 &
                - &
                9.08 &
                - &
                85.20 &
                - &
                84.82 &
                - &
                \textbf{42.94} &
                - &
                \textbf{71.89} &
                - &
                27.09 &
                - &
                78.95 \\
            DOBF &
                - &
                15.45 &
                - &
                75.52 &
                - &
                74.62 &
                - &
                10.28 &
                - &
                87.93 &
                - &
                85.01 &
                - &
                39.35 &
                - &
                69.30 &
                - &
                29.16 &
                - &
                79.10 \\
            T5-learning &
                10 &
                17.62 &
                - &
                77.05 &
                - &
                76.30 &
                3 &
                10.94 &
                - &
                88.46 &
                - &
                86.42 &
                - &
                38.06 &
                - &
                66.58 &
                28 &
                29.90 &
                - &
                78.44 \\
            PLBART &
                19.21 &
                19.40 &
                77.02 &
                78.03 &
                - &
                77.58 &
                8.98 &
                11.05 &
                88.5 &
                88.48 &
                - &
                86.67 &
                - &
                \underline{41.74} &
                - &
                68.42 &
                - &
                33.08 &
                - &
                80.07 \\
            ProphetNet-Code &
                - &
                17.23 &
                - &
                75.40 &
                - &
                75.60 &
                - &
                10.75 &
                - &
                86.82 &
                - &
                84.19 &
                - &
                39.66 &
                - &
                67.24 &
                - &
                29.63 &
                - &
                77.12 \\
            CoTexT &
                21.58 &
                \underline{21.33} &
                77.28 &
                77.20 &
                77.38 &
                77.75 &
                13.03 &
                13.37 &
                88.68 &
                87.13 &
                84.41 &
                85.14 &
                - &
                40.36 &
                - &
                70.16 &
                - &
                31.87 &
                - &
                80.00 \\
            TreeBERT &
                - &
                20.73 &
                - &
                \underline{79.38} &
                - &
                75.17 &
                - &
                12.89 &
                - &
                \underline{89.05} &
                - &
                \textbf{87.15} &
                - &
                41.73 &
                - &
                70.14 &
                - &
                \underline{33.20} &
                - &
                80.46 \\
            CodeT5 &
                21.61 &
                \textbf{21.65} &
                77.43 &
                77.55 &
                - &
                77.24 &
                13.96 &
                \textbf{14.30} &
                87.64 &
                \textbf{89.23} &
                - &
                \underline{87.05} &
                - &
                40.52 &
                - &
                \underline{71.29} &
                - &
                \textbf{34.83} &
                - &
                \textbf{80.75} \\
            SynCoBERT &
                - &
                20.32 &
                - &
                78.81 &
                - &
                78.56 &
                - &
                11.17 &
                - &
                87.94 &
                - &
                86.10 &
                - &
                41.52 &
                - &
                70.26 &
                - &
                29.86 &
                - &
                80.16 \\
            SPT-Code &
                17.54 &
                18.59 &
                75.10 &
                78.51 &
                - &
                74.97 &
                10.86 &
                12.06 &
                87.88 &
                88.37 &
                - &
                84.35 &
                - &
                40.20 &
                - &
                68.97 &
                - &
                33.00 &
                - &
                79.18 \\
            UniXcoder &
                - &
                19.05 &
                - &
                79.18 &
                - &
                \underline{79.45} &
                - &
                \underline{13.96} &
                - &
                87.59 &
                - &
                86.23 &
                - &
                41.69 &
                - &
                69.84 &
                - &
                29.78 &
                - &
                80.02 \\
            \bottomrule
        \end{NiceTabular}
    }
\end{table*}
\begin{table*}[t!]
    \centering
    \caption{Experimental results on code generation tasks involving natural language.}
    \label{table:results_gen_nl}
    \resizebox{\linewidth}{!}{%
        \begin{NiceTabular}{lrrrrrrrrrrrr|rrrrrr}
        \CodeBefore
            \rowcolors{5}{}{gray!15}
        \Body
            \toprule
                \Block{4-1}{\textbf{Model}} &
                \Block{1-12}{\textbf{Code Summarization}} &
                &
                &
                &
                &
                &
                &
                &
                &
                &
                &
                &
                \Block{2-6}{\textbf{Code}\\ \textbf{Generation}} &
                &
                &
                &
                &
                \\
            \cline{2-13}
                &
                \Block[c]{1-2}{\textbf{Java}} &
                &
                \Block[c]{1-2}{\textbf{Py}} &
                &
                \Block[c]{1-2}{\textbf{JS}} &
                &
                \Block[c]{1-2}{\textbf{PHP}} &
                &
                \Block[c]{1-2}{\textbf{Go}} &
                &
                \Block[c]{1-2}{\textbf{Ruby}} &
                &
                &
                &
                &
                &
                &
                \\
            \cline{2-19}
                &
                \Block[c]{1-2}{\textbf{BLEU}} &
                &
                \Block[c]{1-2}{\textbf{BLEU}} &
                &
                \Block[c]{1-2}{\textbf{BLEU}} &
                &
                \Block[c]{1-2}{\textbf{BLEU}} &
                &
                \Block[c]{1-2}{\textbf{BLEU}} &
                &
                \Block[c]{1-2}{\textbf{BLEU}} &
                &
                \Block[c]{1-2}{\textbf{EM}} &
                &
                \Block[c]{1-2}{\textbf{BLEU}} &
                &
                \Block[c]{1-2}{\textbf{CodeBLEU}} &
                \\
            \cline{2-19}
                &
                \textbf{Cur} &
                \textbf{New} &
                \textbf{Cur} &
                \textbf{New} &
                \textbf{Cur} &
                \textbf{New} &
                \textbf{Cur} &
                \textbf{New} &
                \textbf{Cur} &
                \textbf{New} &
                \textbf{Cur} &
                \textbf{New} &
                \textbf{Cur} &
                \textbf{New} &
                \textbf{Cur} &
                \textbf{New} &
                \textbf{Cur} &
                \textbf{New} \\
            \midrule
                Transformer &
                16.26 &
                16.37 &
                15.81 &
                16.47 &
                11.59 &
                10.26 &
                22.12 &
                23.41 &
                16.38 &
                16.46 &
                11.18 &
                11.09 &
                - &
                6.10 &
                - &
                21.67 &
                - &
                26.98 \\
            \hline
                RoBERTa &
                16.47 &
                17.38 &
                18.14 &
                17.49 &
                11.90 &
                11.75 &
                24.02 &
                \underline{24.42} &
                17.72 &
                17.63 &
                11.17 &
                11.26 &
                - &
                \textbf{19.30} &
                - &
                \underline{31.92} &
                - &
                \textbf{35.40} \\
            GPT-2 &
                - &
                18.62 &
                - &
                18.92 &
                - &
                14.13 &
                - &
                23.91 &
                - &
                17.59 &
                - &
                12.03 &
                17.35 &
                17.55 &
                25.37 &
                23.62 &
                29.69 &
                29.93 \\
            BART &
                - &
                \textbf{18.98} &
                - &
                \underline{19.37} &
                - &
                \textbf{14.68} &
                - &
                \textbf{24.46} &
                - &
                \textbf{18.80} &
                - &
                \underline{13.65} &
                - &
                \underline{18.90} &
                - &
                31.38 &
                - &
                \underline{33.72} \\
            T5 &
                18.35 &
                \underline{18.87} &
                19.26 &
                \textbf{19.57} &
                14.57 &
                \underline{14.45} &
                24.59 &
                24.32 &
                19.17 &
                \underline{18.70} &
                14.18 &
                \textbf{13.72} &
                18.65 &
                18.70 &
                32.74 &
                \textbf{32.02} &
                35.95 &
                33.26 \\
            \hline
                CuBERT &
                - &
                16.75 &
                - &
                17.77 &
                - &
                11.34 &
                - &
                22.76 &
                - &
                16.09 &
                - &
                10.46 &
                - &
                19.05 &
                - &
                30.14 &
                - &
                32.82 \\
            GPT-C &
                - &
                17.18 &
                - &
                17.78 &
                - &
                12.01 &
                - &
                23.42 &
                - &
                16.96 &
                - &
                10.54 &
                - &
                \underline{19.85} &
                - &
                30.45 &
                - &
                33.10 \\
            C-BERT &
                - &
                17.44 &
                - &
                18.39 &
                - &
                \underline{13.14} &
                - &
                \underline{23.90} &
                - &
                \underline{17.47} &
                - &
                12.14 &
                - &
                19.80 &
                - &
                33.62 &
                - &
                35.99 \\
            JavaBERT &
                - &
                \underline{18.2}3 &
                - &
                17.57 &
                - &
                11.91 &
                - &
                22.87 &
                - &
                17.13 &
                - &
                10.94 &
                - &
                18.45 &
                - &
                34.62 &
                - &
                36.93 \\
            CodeGPT-adapted &
                - &
                17.68 &
                - &
                \underline{18.46} &
                - &
                12.91 &
                - &
                \textbf{24.68} &
                - &
                17.38 &
                - &
                \underline{12.39} &
                20.10 &
                \textbf{20.15} &
                32.79 &
                \underline{35.94} &
                35.98 &
                \underline{37.27} \\
            DeepDebug &
                - &
                \textbf{19.00} &
                - &
                \textbf{18.85} &
                - &
                \textbf{14.39} &
                - &
                23.37 &
                - &
                \textbf{17.68} &
                - &
                \textbf{13.27} &
                - &
                18.40 &
                - &
                \textbf{36.52} &
                - &
                \textbf{38.90} \\
            \hline
                CodeBERT &
                17.65 &
                18.61 &
                19.06 &
                19.23 &
                14.90 &
                14.75 &
                25.16 &
                24.70 &
                18.07 &
                18.26 &
                12.16 &
                12.53 &
                - &
                21.15 &
                - &
                31.45 &
                - &
                35.26 \\
            GraphCodeBERT &
                - &
                18.93 &
                - &
                19.39 &
                - &
                14.90 &
                - &
                25.64 &
                - &
                18.50 &
                - &
                12.63 &
                - &
                21.00 &
                - &
                34.33 &
                - &
                37.55 \\
            CugLM &
                - &
                18.04 &
                - &
                18.20 &
                - &
                14.07 &
                - &
                24.66 &
                - &
                18.53 &
                - &
                11.47 &
                - &
                21.80 &
                - &
                33.70 &
                - &
                35.91 \\
            DOBF &
                19.05 &
                19.18 &
                18.24 &
                18.41 &
                - &
                13.22 &
                - &
                23.83 &
                - &
                18.28 &
                - &
                13.21 &
                - &
                20.35 &
                - &
                35.26 &
                - &
                37.41 \\
            T5-learning &
                - &
                18.84 &
                - &
                18.23 &
                - &
                13.18 &
                - &
                23.10 &
                - &
                17.29 &
                - &
                12.51 &
                - &
                18.95 &
                - &
                35.76 &
                - &
                38.30 \\
            PLBART &
                18.45 &
                19.31 &
                19.30 &
                19.41 &
                15.56 &
                15.73 &
                23.58 &
                24.47 &
                18.91 &
                19.01 &
                14.11 &
                14.15 &
                18.75 &
                19.85 &
                36.69 &
                36.63 &
                38.52 &
                39.29 \\
            ProphetNet-Code &
                19.39 &
                19.29 &
                17.87 &
                18.20 &
                16.60 &
                \underline{15.95} &
                24.57 &
                24.28 &
                18.43 &
                18.31 &
                14.37 &
                \underline{14.39} &
                - &
                21.70 &
                - &
                37.67 &
                - &
                39.79 \\
            CoTexT &
                19.10 &
                19.19 &
                19.52 &
                \underline{19.72} &
                14.77 &
                15.08 &
                24.47 &
                24.57 &
                19.37 &
                19.13 &
                13.07 &
                14.28 &
                20.10 &
                21.80 &
                36.51 &
                \underline{38.74} &
                39.49 &
                40.63 \\
            TreeBERT &
                - &
                18.90 &
                - &
                19.44 &
                - &
                15.05 &
                - &
                23.82 &
                - &
                18.74 &
                - &
                13.57 &
                - &
                22.05 &
                - &
                38.27 &
                - &
                39.73 \\
            CodeT5 &
                20.31 &
                \textbf{20.35} &
                20.01 &
                \textbf{20.17} &
                16.16 &
                \textbf{16.75} &
                26.03 &
                \textbf{25.97} &
                19.56 &
                \textbf{19.68} &
                15.24 &
                \textbf{15.36} &
                22.30 &
                \textbf{23.40} &
                40.73 &
                \textbf{40.75} &
                43.20 &
                \textbf{43.40} \\
            SynCoBERT &
                - &
                18.89 &
                - &
                18.74 &
                - &
                14.57 &
                - &
                25.55 &
                - &
                18.36 &
                - &
                13.91 &
                - &
                21.35 &
                - &
                38.39 &
                - &
                \underline{41.21} \\
            SPT-Code &
                - &
                18.90 &
                - &
                19.71 &
                - &
                15.28 &
                - &
                24.78 &
                - &
                \underline{19.17} &
                - &
                14.38 &
                - &
                20.00 &
                - &
                37.91 &
                - &
                39.90 \\
            UniXcoder &
                - &
                \underline{19.42} &
                - &
                18.64 &
                - &
                14.27 &
                - &
                \underline{25.70} &
                - &
                18.59 &
                - &
                14.32 &
                22.60 &
                \underline{22.65} &
                - &
                38.73 &
                - &
                40.86 \\
            \bottomrule
        \end{NiceTabular}
    }
\end{table*}

\subsection{Code Question Answering}

The new SOTA model remains the same as the current one, i.e., UniXcoder. But our newly reported SOTA performance has an improvement of 0.2 percent MRR points. Note that \textit{MCL}, the pre-training task used by the SOTA model UniXcoder, aims to distinguish whether two inputs match each other, which is also the goal of the Code Question Answering task. 

\subsection{Code Translation}

Although the models are ranked according to their average EM value on the ``Java to C\#'' and the ``C\# to Java'' sub-datasets, we find that the Top-2 models on the two sub-datasets are both PLBART and CodeT5\footnote{For a discussion of the results w.r.t.\ other evaluation metrics. See the supplementary file.}. Besides, the Top 3--6 models on this task are CoTexT, SPT-Code, DOBF, and UniXcoder respectively. 

\subsubsection{Architecture} According to the new results, that \textbf{TF-based models take the absolute lead on this task} can be verified, since the Top-5 models are all TF-based, and given that we have more TF-based models in our comparison than before, the rank of the best performing TE-based model (i.e., SynCoBERT in Current and UniXcoder in New) drops from fourth to sixth.

\subsubsection{Modality} \textbf{The importance of NL is well validated}, due to the fact that the Top-4 performers in both the current (i.e., CodeT5, PLBART, SPT-Code and SynCoBERT) and the new results (i.e., PLBART, CodeT5, CoTexT and SPT-Code) use NL. The role of code structure, on the other hand, is less clear, since the Top-2 models (i.e., PLBART, CodeT5) are not pre-trained on the Structure modality and the best model using code structure drops from the third place in ``Cur'' (i.e., SynCoBERT)  to the fourth place in ``New'' (i.e., SPT-Code). 

\subsubsection{Pre-training Tasks} The new results show that \textbf{the pre-training objective \textit{DAE} has a more significant impact than \textit{BDG/CMG}}. To exemplify, consider PLBART and CodeT5, both of which are TF-based and employ the same modalities (code and NL). They differ only in terms of the pre-training tasks: the former uses \textit{DAE} and the latter uses \textit{BDG/CMG}. The fact that PLBART outperforms CodeT5 can therefore be attributed to the fact that \textit{DAE} is a better pre-training task for Code Translation than \textit{BDG/CMG}. This conclusion is contrary to the conclusion drawn from the CUR results, where \textit{BDG} is believed to have a stronger influence than \textit{DAE} on Code Translation due to the fact that CodeT5 beat PLBART by 1.3 percent EM points.

\subsection{Bug Fixing}

Considering the EM value averaged over the ``small'' and ``medium'' datasets, the Top-4 models change from CodeT5, CoTexT, DeepDebug, and SPT-Code (listed in decreasing order of performance) to CodeT5, CoTexT, TreeBERT, and UniXcoder. A closer examination of the sub-datasets reveals that UniXcoder outperforms CoText and TreeBERT, achieving the second best performance on the ``medium'' dataset while ranking 4th on the ``small'' one. 

\subsubsection{Architecture} The Top-3 performance is achieved by three TF-based models (i.e, CodeT5, CoTexT, and TreeBERT) and the best and second TE-based models (i.e., UniXcoder and SynCoBERT) rank 4th and 5th respectively. Besides, the best TD-based model (i.e., CodeGPT-adapted) only rank 14th. This seems to suggest that \textbf{the TF architecture should be considered first when designing high performance pre-trained models for this task}.

\subsubsection{Pre-training Tasks} \textbf{The most useful pre-training tasks for Bug Fixing is the sequence-to-sequence variants of MLM} adapted to the Transformer decoder. They enable a model to acquire the ability to generate target sequences from an incomplete one. As an example, consider the top-3 TF-based models, which all use such pre-training tasks: \textit{Seq2seq MLM} in CodeT5 and CoTexT, \textit{TMLM} in TreeBERT, and \textit{Seq2seq IMLM} in CodeT5. Moreover, by using Seq2seqMLM as the only pre-training task, the second-highest ranked model, CoTexT, achieves better performance than TreeBERT, which uses \textit{NOP} in addition to \textit{TMLM} for pre-training. 

\subsection{Code Completion}

This is the only SE task among the ones we consider where SOTA performance is achieved by the TD-based model CodeGPT-adapted, and it is the SOTA model in both the current and new results. This seems to suggest the absolute dominance of the TD architecture on this task. Our new results further suggest that \textbf{the pre-training objective \textit{ULM} (adopted by the Top-3 models on this task, i.e., CodeGPT-adapted, CugLM, and GPT-C), whose goal is similar to that of code completion, plays an influential role in Code Completion}. As an example, consider the TE-based model CugLM, which outperforms another TE-based model CuBERT (pretrained on \textit{MLM} and \textit{NSP}) and achieves the second best performance by using \textit{ULM} in addition to \textit{MLM} and \textit{NSP}. Moreover, in terms of modality, \textbf{Code Completion is the only task where neither NL nor code structure plays a positive role} since all of the Top-3 models use code as the only input modality. We speculate the reason is that there is currently no effective way to combine these two modalities with \textit{ULM}.

\subsection{Assert Generation}

Since only T5-learning has been evaluated on this task currently, all conclusions drawn from the new results could be viewed as new findings. First, \textbf{the Top-5 performers are all TF-based models} (i.e., PLBART, TreeBERT, SPT-Code, T5-learning, and CodeT5 by order). The best performing TE-based (i.e., UniXcoder) and TD-based (i.e., CodeGPT-adapted) models rank 8th and 16th, respectively. As far as modality is concerned, NL seems to have a greater impact than other modalities, as four of the Top-5 models (i.e., PLBART, SPT-Code, T5-learning, and CodeT5)  use NL, whereas only two (i.e., TreeBERT and SPT-Code) use code structures. 

\subsection{Mutant Generation}

\textbf{NL and code structure appear to have positive impacts}, since the Top-3 models (i.e., CodeT5, TreeBERT, and PLBART by order) use either NL or code structure as one of the inputs in addition to the code. As for pre-training tasks, \textit{DAE} alone is able to help the model (i.e., PLBART) achieve the third best performance. The structure-aware pre-training tasks, such as \textit{TMLM} and \textit{NOP} used by TreeBERT (the second best) and \textit{CAP} used by SPT-Code (the fourth best) clearly have  positive impacts on this task.  

\subsection{Code Summarization}

\subsubsection{Architecture} The best TE-based model (UniXcoder) ranks 5th, with the Top-4 being TF-based models (i.e., CodeT5, ProphetNet-Code, CoTexT, and SPT-Code), which suggests \textbf{the strong positive influence of the TF architecture on this task}. This is not in line with the current results in which the best TE-based model ranked second. With the new results, the best TD-based model (GPT-2) ranks 15th. 

\subsubsection{Modality} \textbf{The highest ranks achieved by models pre-trained on NL only (e.g., T5 and BART) are 5th and 9th in the current and new results, respectively}. The reasons why they perform even better than some of the models pre-trained on code or structure in addition to NL (e.g., CodeBERT, GraphCodeBERT, etc.) are two-fold. First, they acquire the ability to generate NL during pre-training, which is required by Code Summarization, and (2) because of the ``naturalness'' of the source code~\cite{ernst2017natural,buratti2020c-bert}, they are able to understand the code to some extent although they only have the ability to understand NL.

\subsubsection{Pre-training Tasks} \textbf{Cross-modal(Code and Natural Language)-aware generation tasks such as \textit{BDG/CMG} and \textit{MNG} have positive impacts on a model's performance on this task}. As an example, CodeT5, which utilizes \textit{BDG}, and SPT-Code, which utilizes \textit{MNG}, are the top performers among TF-based models, and UniXcoder, which utilizes \textit{CMG}, is the top performer among TE-based models. 

\subsection{Code Generation}

The SOTA model changes from TE-based (UniXcoder) to TF-based (CodeT5), and the best TE-based (UniXcoder) and TD-based (CodeGPT-adapted) models rank 2nd and 11th, respectively. While the new SOTA model (i.e., CodeT5) for Code Generation is also the SOTA model for Code Summarization, \textbf{the ranks of T5 and BART (pre-trained only on NL) on this task are lower than their ranks on Code Summarization}, because understanding code and generating code are fundamentally different in nature. In addition, \textbf{the importance of the code and NL modalities for this task is not as clear as that for Code Summarization}, considering that among the Top-3 models (CodeT5, UniXcoder, and TreeBERT), only CodeT5 uses both code and NL: UniXcoder uses code structure and NL and TreeBERT uses code structure and code instead. Moreover, although only NL is the input of this task, \textbf{pre-training on code structure has a positive impact on this task}, since both of the second and third best models (UniXcoder and TreeBERT) are pre-trained on tasks such as \textit{MCL}, \textit{CMG}, and \textit{NOP}.

\section{Insights and Takeaways}
\label{section:insights}

After analysis and discussion by task, we have some insights and takeaways to provide to subsequent researchers.

\begin{itemize}
    \item When designing a new model to solve multiple tasks, look up the current SOTA model's architecture, features, and pre-training tasks for each task, and use such information as a starting point.
    \item Always pre-train on multiple programming languages.
    \item Always pre-train with NL, since all of the new SOTAs use NL.
    \item Utilize structure information in PTMs for code understanding tasks.
    \item Ensure the pre-training tasks are as similar in form as possible to the target downstream task.
    \item Use different CodePTMs for different target task types since there is no almighty CodePTM, as per our results and Zeng et al.~\cite{zeng2022extensive}.
\end{itemize}

Particularly, for the following tasks, we have additional takeaways:

\begin{itemize}
    \item \textbf{Clone Detection}: Although the TE-based model achieves the best performance, comparable results are achieved with the TF-based model. Besides, the use of NL and code structure is beneficial. Finally, MLM and its variants have better results on this task.
    \item \textbf{Code-to-Code Retrieval}: Utilize NL and code structure following the "Altogether" strategy. Besides, MLM and its variants, as well as structure-aware pre-training tasks, have positive effects on this task.
    \item \textbf{Code Question Answering}: Prefer TE models and use NL whenever possible.
    \item \textbf{Assert Generation}: NL is not a required modality. The reason is that although the model with the best performance uses NL, NL is not used in the same data sample as other modalities (because of the Standalone strategy). Seq2seq pre-training tasks, such as DAE, MASS and MNG, should be prioritized.
    \item \textbf{Code Generation}: Besides TF, TE is worth trying. The use of NL-Code generation code pre-training tasks (e.g., BDG and CMG) is mandatory.
\end{itemize}

Finally, through our experiments, we propose several possible subsequent research directions as follows. 

\begin{itemize}
    \item Design more efficient pre-training tasks to make CodePTMs learn source code features better~\cite{karmakar2021pre}.
    \item Improve the efficiency of CodePTMs for fine-tuning on downstream tasks~\cite{wang2022bridging}.
    \item Make the large CodePTMs lighter~\cite{zhang2022diet,shi2022compressing}.
    \item Improve the robustness of CodePTMs.
\end{itemize}

\section{Threats to Validity}
\label{section:threats}

\subsubsection*{Construct Validity}

As discussed in Section~\ref{section:setup_implementation}, we have re-implemented some PTMs (category IV) or re-collected some datasets (category III and some in IV). The replication may not be perfect but we have tried our best to do the re-implementation and collect the datasets to minimize the deviations from the original model (See Section~\ref{section:setup_implementation}). Besides, we adopt the statistical significance testing to measuring the differences between our implementation and the original ones. 

\subsubsection*{Internal Validity}

It is widely agreed that, during fine-tuning, hyperparameters have a significant impact on the performance of pre-trained models. For models where hyperparameters for fine-tuning are not available (See Section~\ref{section:setup_implementation}), the settings we obtain by hyperparameter search may introduce some bias with the performance reported in the original paper. But we have tried best to derive best performance of these models on each SE task. 

\subsubsection*{External Validity}

The results and observations we obtained in this work may apply only to the downstream tasks and corresponding datasets we have evaluated. For the other SE tasks and datasets, we cannot guarantee exactly the same results and observations.

\section{Conclusion}
\label{section:conclusion}

We conducted the first systematic empirical comparison of existing pre-trained models of source code\footnote{All materials used in our experiments are available at \url{https://github.com/NougatCA/FineTuner} and \url{https://doi.org/10.5281/zenodo.7318110}.}. We believe that the results of our large-scale evaluation and the associated discussion can provide SE researchers with a better understanding of existing PTMs and their relative strengths and weaknesses, as well as a better characterization of the state-of-the-art of each SE task on which PTMs are commonly evaluated.

This paper provides many valuable findings that are either not available based on the existing results alone or completely contrary to current findings. For example, we found that TF-based models have clear advantages for not only code generation tasks but also code understanding tasks. We hope that this paper could provide interested researchers with a comprehensive and comparable insights into the current state of this domain and inspire them to design more powerful pre-trained models of source code.

\section*{Acknowledgment}
This work was supported by National Natural Science Foundation of China (61802167), Cooperation Fund of Huawei-NJU Creative Laboratory for the Next Programming, and NSF award 2034508. We also thank the reviewers for their helpful comments. Chuanyi Li is the corresponding author.
\bibliographystyle{IEEEtran}
\bibliography{refs}

\end{document}